\documentclass[a4paper,11pt]{article}
\pdfoutput=1 % if your are submitting a pdflatex (i.e. if you have
\usepackage{jheppub} % for details on the use of the package, please
\usepackage{amsmath,amssymb,amsthm,amscd,graphicx}
\input epsf.sty

\theoremstyle{definition}

\newcommand{\bs}{\boldsymbol}

\newcommand{\CC}{{\cal C}}
\newcommand{\CE}{{\cal E}}
\newcommand{\CF}{{\cal F}}
\newcommand{\CG}{{\cal G}}
\newcommand{\CH}{{\cal H}}
\newcommand{\CI}{{\cal I}}
\newcommand{\CJ}{{\cal J}}

\newcommand{\CL}{{\cal L}}
\newcommand{\CM}{{\cal M}}
\newcommand{\CN}{{\cal N}}
\newcommand{\CO}{{\cal O}}

\newcommand{\CR}{{\cal R}}
\newcommand{\CS}{{\cal S}}

\def\IZ{{\mathbb Z}}
\def\IR{{\mathbb R}}

\newcommand{\re}{{\rm e}}
\newcommand{\ri}{{\rm i}}
\newcommand{\rd}{{\rm d}}

\newcommand{\mfp}{\mathfrak{p}}

\newcommand{\be}{\begin{equation}}
\newcommand{\ee}{\end{equation}}
\newcommand{\ba}{\begin{aligned}}
\newcommand{\ea}{\end{aligned}}
\newcommand{\ben}{\begin{eqnarray}\displaystyle}
\newcommand{\een}{\end{eqnarray}}

\newcommand{\sectiono}[1]{\section{#1}\setcounter{equation}{0}}

\newdimen\tableauside\tableauside=1.0ex
\newdimen\tableaurule\tableaurule=0.4pt
\newdimen\tableaustep
\def\phantomhrule#1{\hbox{\vbox to0pt{\hrule height\tableaurule width#1\vss}}}
\def\phantomvrule#1{\vbox{\hbox to0pt{\vrule width\tableaurule height#1\hss}}}
\def\sqr{\vbox{%
  \phantomhrule\tableaustep
  \hbox{\phantomvrule\tableaustep\kern\tableaustep\phantomvrule\tableaustep}%
  \hbox{\vbox{\phantomhrule\tableauside}\kern-\tableaurule}}}
\def\squares#1{\hbox{\count0=#1\noindent\loop\sqr
  \advance\count0 by-1 \ifnum\count0>0\repeat}}
\def\tableau#1{\vcenter{\offinterlineskip
  \tableaustep=\tableauside\advance\tableaustep by-\tableaurule
  \kern\normallineskip\hbox
    {\kern\normallineskip\vbox
      {\gettableau#1 0 }%
     \kern\normallineskip\kern\tableaurule}%
  \kern\normallineskip\kern\tableaurule}}
\def\gettableau#1{\ifnum#1=0\let\next=\null\else
\squares{#1}\let\next=\gettableau\fi\next}

\def\bPhi{\boldsymbol{\Phi}}

\newcommand{\figref}[1]{Fig.~\protect\ref{#1}}
\title{\Huge{\boldmath Anatomy of the simplest renormalon}}

\author{Marcos Mari\~no}

\affiliation{D\'epartement de Physique Th\'eorique et Section de Math\'ematiques\\
Universit\'e de Gen\`eve, Gen\`eve, CH-1211 Switzerland}

\emailAdd{Marcos.Marino@unige.ch} 

\abstract{Perhaps the simplest IR renormalon occurs in the ground state energy of a superrenormalizable model, the scalar 
$O(N)$ theory in two dimensions with a quartic potential and negative squared mass. We show that this renormalon, found previously in 
perturbation theory at next-to-leading order (NLO) in the $1/N$ expansion, gives indeed the correct asymptotic expansion 
of the exact large $N$ solution of the model, and we determine explicitly the complete trans-series of non-perturbative 
corrections to the perturbative result. We also use this framework to study the $O(N)$-invariant two-point function of the 
scalar field. As expected, it is IR finite in perturbation theory, but it is afflicted as well with an IR renormalon singularity and is not Borel summable. The pole 
mass is purely non-perturbative and its trans-series can be also fully determined at NLO in $1/N$. }    

\begin{document}
\maketitle
\flushbottom
 
\sectiono{Introduction}

Factorial divergence due to renormalons is a generic feature of the perturbative approach to 
quantum field theory (QFT). Mathematically, 
renormalons appear as singularities of the Borel transform of the perturbative series. Since their inception in the late 1970s, they have been 
interpreted as smoking guns for nonperturbative effects which do not have a semiclassical origin. 

Renormalons have been mostly studied in renormalizable, asymptotically free theories. 
When they arise in the correlation function of products of operators, their presence has been related \cite{parisi-renormalons} to 
the existence of non-trivial condensates in the operator product expansion (OPE) \cite{svz}. However, renormalons also appear in more 
general observables without an OPE description. A particularly simple example was discovered in \cite{mr-new}, 
in the ground state energy of the two-dimensional scalar field theory with a $O(N)$ symmetry, a quartic potential, 
and a negative squared mass. This theory is super-renormalizable, but it is well-known that its IR physics is subtle. This is because 
the $O(N)$ symmetry cannot be spontaneously broken in two dimensions, due to the Coleman--Mermin--Wagner 
theorem \cite{coleman2d,mermin-wagner}. However, as explained by Jevicki a long time ago \cite{jevicki}, the conventional 
perturbation theory in this model can be obtained by expanding around a ``false" vacuum in which the $O(N)$ 
symmetry is broken, and the resulting perturbative series for the ground state energy is free of IR divergences. 
It was shown in \cite{mr-new} that this series is factorially divergent and non-Borel summable due to an 
IR renormalon, by working at next-to-leading order (NLO) in the $1/N$ expansion, 
but perturbatively in the coupling constant, as usual in renormalon analysis \cite{beneke}. This behaviour is in stark contrast 
to the quartic scalar theory with a positive squared mass, and to the $N=1$ theory 
with a negative mass, where in both cases the perturbative series is Borel summable \cite{eckmann,serone1, serone2, serone3}. 
In the theory with a positive squared mass, the ground state energy at NLO in $1/N$ is 
even convergent \cite{marco}. 
For this reason, the non-Borel summability found in the $O(N)$ theory was interpreted 
in \cite{mr-new} as a manifestation of the absence of Goldstone bosons in two dimensions. 

It is perhaps useful to compare this situation with the double-well potential 
in quantum mechanics. In that case, the $\IZ_2$ symmetry 
can not be spontaneously broken, yet we do perturbation theory by expanding around one of the 
two ``false" vacua. The resulting 
perturbative series is factorially divergent and non-Borel summable, but we can cure the problem 
by taking into account the non-perturbative effects due to instantons. When these effects are 
added, in the form of a trans-series, one can reproduce 
the exact ground state energy obtained from the spectral theory of the Schr\"odinger operator 
(see e.g. \cite{mm-advanced} for 
an exposition and references).  

 \begin{figure}[!ht]
\leavevmode
\begin{center}
\includegraphics[height=3cm]{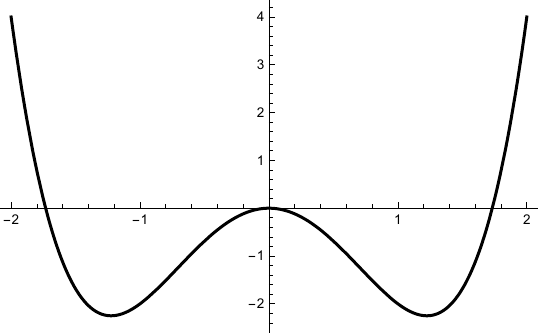}\qquad \qquad \qquad \includegraphics[height=4.5cm]{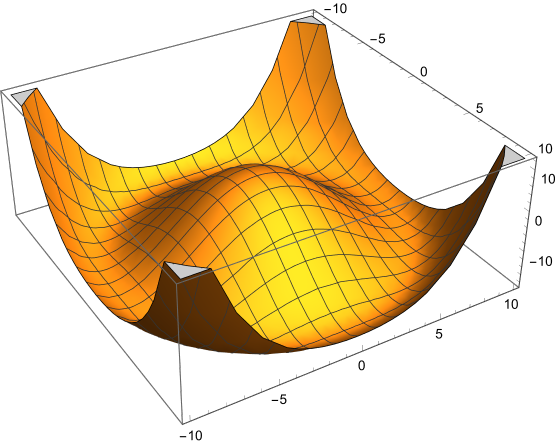}
\end{center}
\caption{In the double-well potential in quantum mechanics (left), the two perturbative vacua 
are false vacua, and expanding around them leads to a non-Borel summable perturbative series. In the 
scalar field theory with a quartic Mexican hat potential and $O(N\ge 2)$ symmetry (right), the perturbative 
vacua are also false vacua in $d=2$, and they also lead to a non-Borel summable perturbative expansion. In the 
quantum mechanical case, the relevant non-perturbative effects are instantons, while in the two-dimensional field theory, they are IR renormalons.}
\label{f-vacua}
\end{figure} 

In the case of our two-dimensional QFT, one can find exact results for the observables, order by order in 
the $1/N$ expansion, by using non-perturbative large $N$ techniques. In contrast to the perturbative large $N$ 
methods based on summing up ``bubble" diagrams, their non-perturbative counterpart produces actual functions 
of the 't Hooft coupling. This non-perturbative approach was used long ago in \cite{cjp} to calculate the effective potential at 
large $N$, and to show that there is no spontaneous symmetry breaking at large $N$. A more detailed large $N$ 
analysis \cite{marco} shows that there is a non-perturbative mass gap, and one can compute various physical 
quantities at NLO in $1/N$, like the ground state energy. 

The perturbative and non-perturbative large $N$ results are obtained by expanding around different 
large $N$ vacua, and they are very different, even as mathematical entities. The first one is a formal 
power series in the 't Hooft coupling, 
while the second one is a well-defined, real function of that coupling. As in quantum mechanics, 
we expect that the formal series gives the asymptotic expansion of the well-defined function, at weak coupling. 
We also expect that the exact result can be obtained by adding a series of 
non-perturbative corrections to the perturbative series, 
and performing an appropriate resummation in which renormalon singularities cancel. In this 
paper we verify these expectations in detail. Mathematically, we ``decode" the exact, non-perturbative answer as a trans-series, and we verify in particular that the perturbative series found in \cite{mr-new} is indeed the correct asymptotic expansion of the non-perturbative large $N$ result. 

This sort of decoding at large $N$ is not new and has been studied in other 
two-dimensional models, starting with the beautiful paper \cite{bbk} on the 
two-point function of the non-linear sigma model, which builds on the 
pioneering work of \cite{david2}. Similar analyses have been done recently for the 
supersymmetric sigma model \cite{sss} 
and the Gross--Neveu model \cite{mm-trans,liu-gn}. We believe that the example of the $O(N)$ quartic theory is 
interesting for various reasons. First of all, it is in many ways technically simpler than previous studies, 
and provides perhaps the simplest example of a QFT renormalon and its associated 
trans-series. Second, it concerns a quantity, the ground state energy, in which the trans-series 
cannot be calculated in principle with the OPE approach, and might serve as a testing ground 
for more general techniques. Finally, this example illustrates an IR sensitive dependence 
whose underlying physics is different from the one in asymptotically free theories. In particular, 
since there is no dimensional transmutation, the ground state energy has a non-trivial dependence 
on the coupling constant, both at the perturbative and the non-perturbative level. 

We also calculate the $O(N)$-invariant two-point function of the scalar field at NLO in 
$1/N$, both in perturbation theory and in the exact $1/N$ expansion. We show explicitly that, in the perturbative calculation, 
IR divergences cancel when we sum all diagrams, as expected from the 
study of the non-linear sigma model \cite{elitzur, david-cancel}. As a result, we find a 
closed form, integral expression for the 
all-loop perturbative series. It displays as well an IR renormalon behavior, with the same set 
of Borel singularities as the ground state energy. In particular, it is not Borel summable, in contrast 
to what happens in e.g. the $N=1$ theory with positive squared mass. Our calculation also shows 
that the pole mass vanishes at all orders in perturbation theory, at NLO in $1/N$, but can be calculated 
as a non-trivial trans-series 
in the non-perturbative large $N$ vacuum. 

The paper is organized as follows. In section \ref{pert-subsec} we focus on the perturbative analysis of the observables and the 
renormalon singularities. We review and slightly extend the results of \cite{mr-new} and we present a detailed computation of the two-point function of the 
$\Phi$ field at NLO in $1/N$ and at all loops, and its renormalon behavior. In section \ref{nonpert-sec} we review the standard large $N$ non-perturbative calculations of the quantities considered in section \ref{pert-subsec}, and we present complete results for the trans-series corrections in the case of the ground state energy and the pole mass. We conclude with some observations and open problems. The Appendix contains technical details 
complementing the computations performed in the bulk of the paper. 

\sectiono{Perturbative series and renormalons}
\label{pert-subsec}

\subsection{The $O(N)$ theory}
In this section we recall some basic facts about the $O(N)$, $\lambda \phi^4$ theory. Further details can be found in e.g. \cite{mr-new}. The theory is described by the Euclidean Lagrangian
\be
\label{lag}
\CL (\bPhi)= {1\over 2} \partial_\mu \bs{\Phi}\cdot \partial^\mu \bs{\Phi}+ V(\bs{\Phi})
\ee
where the scalar potential is
\be
\label{spot}
V(\bs{\Phi})= {\mu^2 \over 2} \bs{\Phi}^2 +{g\over 4!} \bs{\Phi}^4. 
\ee
We will consider the stable case $g>0$, and the classically ordered phase with
\be
\mu^2 <0. 
\ee
In this case, the classical minimum occurs at 
\be
\label{classm}
\bs{\Phi}^2 :=\phi^2= - 3! {\mu^2 \over g}. 
\ee
We will take the vacuum configuration of the field to be in the direction 
\be
\bs{\Phi}_0 = \left( \phi, 0, \cdots, 0 \right). 
\ee
In this classical vacuum, the $O(N)$ symmetry is spontaneously broken down to $O(N-1)$. There are $N-1$ 
Goldstone bosons and one massive particle, with mass
\be
\label{mass}
m^2= -2 \mu^2.
\ee
The quantum theory is superrenormalizable. As is well-known, scalar field theories in two dimensions 
can be renormalized by normal-ordering the fields. This is equivalent to adding mass counterterms and vacuum energy counterterms, and different choices of scheme can be parametrized by the mass involved 
in the counterterm. The simplest choice is the one made in \cite{jevicki}, in which this mass is simply (\ref{mass}). 
This is equivalent to renormalize 
\be
\label{reg-int}
\int  {\rd^2 k \over (2 \pi)^2}{1\over k^2+a^2}= -{1\over 4 \pi} \log \left( {a^2 \over m^2} \right), 
\ee
as well as
  \be
  \label{log-reg}
  \int {\rd^2 k \over (2 \pi)^2} \log(k^2+ a^2) =    { a^2 \over 4 \pi}  \left(1- \log \left( {a^2 \over m^2} \right) \right).
  \ee

\subsection{The perturbative large $N$  vacuum}

We will now review and extend some results of \cite{mr-new} on the calculation 
of the ground state energy, at NLO in the $1/N$ expansion but perturbatively in the 't Hooft parameter
\be
\kappa= {g N \over 3}. 
\ee
To work at large $N$ we follow \cite{cjp} and perform a Hubbard--Stratonovich transformation. In this way we obtain 
an equivalent theory of two fields, $\bPhi$ and $X$, with Lagrangian, 
\be
\ba
\label{lag-chi}
\CL (\bPhi, X)&= \CL(\bPhi) -{N \over 2 \kappa} \left(X- {\kappa \over 2 N} \bPhi^2 -\mu^2\right)^2\\
&= {1\over 2} \partial_\mu \bPhi \cdot \partial^\mu \bPhi +{X \over 2} \bPhi^2 - { N \over 2 \kappa} X^2 +{N \mu^2 \over \kappa} X - {N  \mu^4 \over 2 \kappa}. 
\ea
\ee
To perform perturbative calculations in this theory we expand around a constant configuration of the fields:
\be
\Phi_1(x) = \xi (x)+{\sqrt{N}} \phi, \qquad X(x)= \chi + \tilde \chi (x), \qquad \Phi_j(x)= \eta_{j-1}(x), \qquad j=2, \cdots, N. 
\ee
When $\phi \not=0$, the $O(N)$ symmetry is broken spontaneously. 
After plugging these expansions in (\ref{lag-chi}) and rescaling $\tilde \chi \rightarrow \tilde \chi/{\sqrt{N}}$, we obtain.
\be
\label{shifted-lag-e}
\ba
\CL  (\bs{\eta},  \xi, \tilde \chi; \phi, \chi)&=V_{\rm tree} (\phi, \chi) +  {1\over 2} \left( \nabla \bs{\eta} \right)^2+{\chi \over 2}  \bs{\eta}^2 +{1\over 2}  \left( \nabla\xi \right)^2+{\chi \over 2} \xi^2 - {\tilde \chi^2 \over2 \kappa} + \tilde \chi \xi \phi  \\
&+{ {\sqrt{N}}  \tilde \chi \over 2}  \left(  \phi^2 - { 2 \chi  \over  \kappa} +{ 2\mu^2\over \kappa} \right) +\sqrt{N} \phi \chi \xi+{1\over 2 {\sqrt{N}}} \tilde \chi \xi^2 +{1\over 2 {\sqrt{N}}}  \tilde \chi \bs{\eta}^2.
\ea
\ee
In this expression, 
\be
V_{\rm tree} (\phi, \chi)= N \left( -{ \chi^2 \over 2 \kappa} +{\chi \phi^2 \over 2} +{ \mu^2 \chi \over \kappa} -{ \mu^4 \over 2 \kappa} \right) 
\ee
is the effective potential at tree level.

 \begin{figure}[!ht]
\leavevmode
\begin{center}
\includegraphics[height=3.5cm]{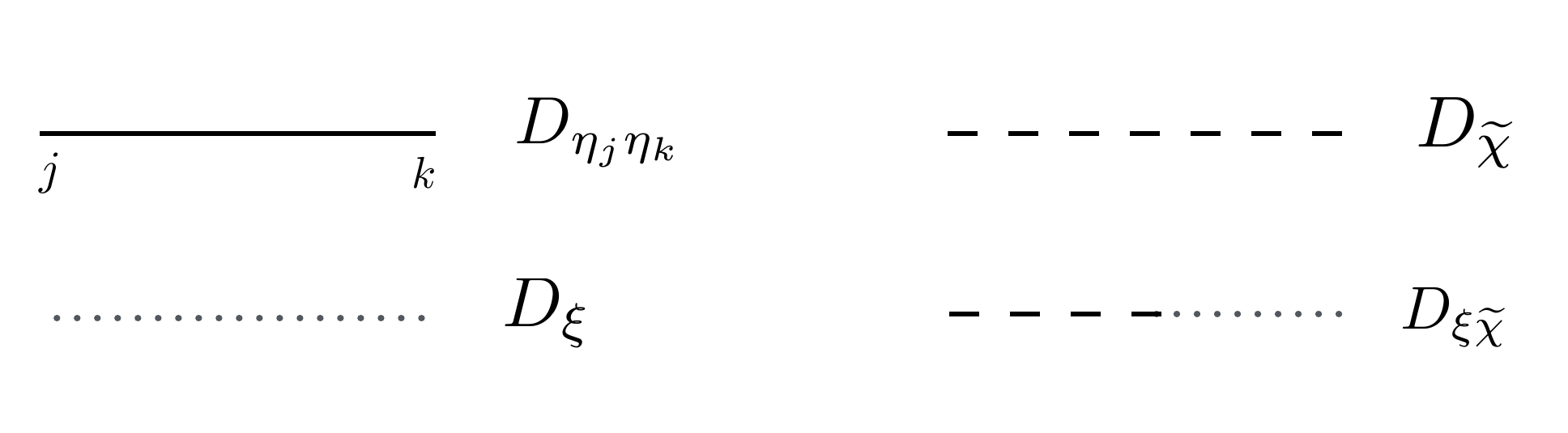}
\end{center}
\caption{The propagators for the effective Lagrangian (\ref{shifted-lag-e}).}
\label{f-eprop}
\end{figure}

\begin{figure}[!ht]
\leavevmode
\begin{center}
\includegraphics[height=3.5cm]{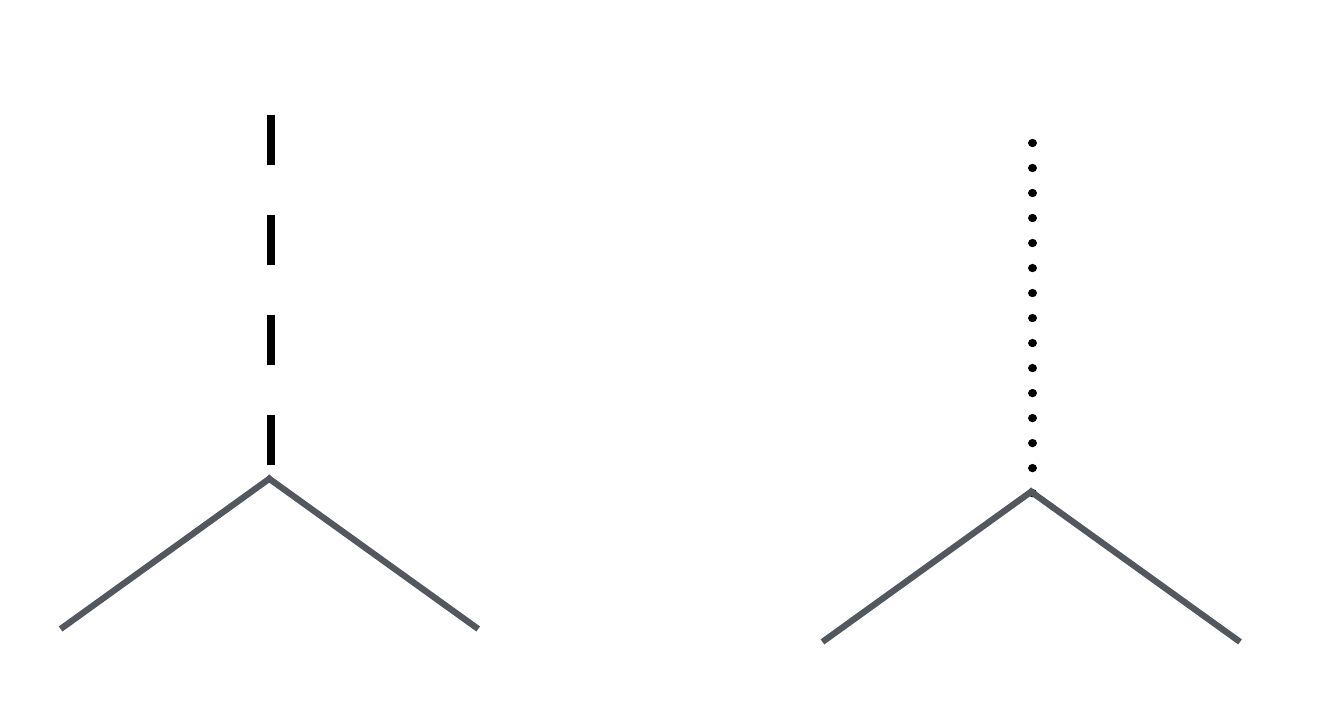}
\end{center}
\caption{The vertices for the effective Lagrangian (\ref{shifted-lag-e}). The first one represents the $\tilde \chi \xi^2$ coupling, while the second 
one represents the $\tilde \chi \eta^2$ coupling.}
\label{f-evertex}
\end{figure}

The Lagrangian (\ref{shifted-lag-e}) involves the fields $\bs{\eta}$, $\tilde \chi$ and $\xi$, with a non-diagonal 
coupling between $\xi$ and $\tilde\chi$. The propagators are given by 
\be
\ba
D_{\eta_j\eta_k}&= {\delta_{jk} \over p^2+ \chi}, \qquad D_{\xi }= {1\over p^2 + \chi + \kappa \phi^2 }, \\
D_{\xi \tilde \chi}&=   {\kappa \phi \over  p^2 + \chi +\kappa \phi^2 }, \qquad
D_{\tilde\chi }= -\kappa {p^2 + \chi \over  p^2 + \chi +\kappa \phi^2 }. 
\ea
\ee
They are shown diagrammatically in \figref{f-eprop}. The $\eta$, $\tilde \chi$ and $\xi$ fields are 
represented by a full line, a dashed line and a point line, respectively. The fields interact 
through two types of cubic vertices shown in \figref{f-evertex}. 

 \begin{figure}[!ht]
\leavevmode
\begin{center}
\includegraphics[height=2.75cm]{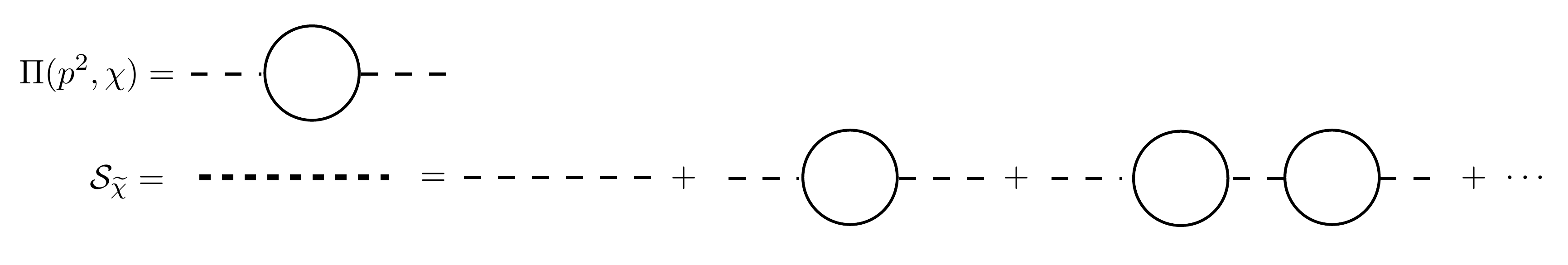}
\end{center}
\caption{The polarization loop gives the function $\Pi(p^2, \chi)$ introduced in (\ref{Pif}). The effective large $N$ propagator for 
the $\tilde \chi$ field, $S_{\tilde \chi}$, is obtained 
by summing up the bubbles.}
\label{f-pol}
\end{figure} 

In this theory there are two fundamental ingredients for large $N$ calculations. The first one is the polarization loop for the $\tilde \chi$ field, given by 
\be
\label{Pif}
\Pi(p^2, \chi)= {1\over 2} \int  {\rd^2 q \over (2 \pi)^2} {1 \over (q^2 + \chi) ( (p+q)^2 +\chi)}={1\over 4 \pi {\sqrt{ p^2(p^2+ 4 \chi)}}} \log { {\sqrt{p^2 + 4\chi}} +{\sqrt{p^2}} \over   {\sqrt{p^2 + 4\chi}} -{\sqrt{p^2}}}. 
\ee
The second ingredient is the effective propagator for the $\tilde \chi$ field, which is obtained by adding up all the bubbles made up of polarization loops. 
This is represented graphically in \figref{f-pol}, and has the expression 
\be
\CS_{\tilde \chi}= {D_{\tilde \chi} \over 1- D_{\tilde \chi} \Pi(p,\chi)}. 
\ee
As we will see in a moment, we are particularly interested in the limit in which $\chi \rightarrow 0$. By using that 
\be
\label{propa-limit}
\Pi(p^2, \chi) = {1\over 4 \pi p^2} \log\left({p^2 \over \chi} \right) + \CO(\chi)
\ee
one finds, as $\chi \to 0$, 
\be
\label{Schi-lim}
 \CS_{\tilde \chi}= - \kappa {p^2 \over   p^2 + m^2 +  m^2 \gamma \log \left( {p^2 \over m^2} \right) }+ \CO(\chi), 
\ee
where we have introduced the dimensionless coupling 
\be
\label{gam-def}
\gamma= {1\over 4 \pi} {\kappa \over m^2}. 
\ee

Let us now briefly review the calculation of the effective potential. It has the $1/N$ expansion
\be
V(\phi, \chi) = N \sum_{k \ge0} V_{(k)} (\phi, \chi) N^{-k}. 
\ee
The leading term in this expansion, $V_{(0)} (\phi, \chi)$, is given by the classical, tree-level potential, plus the one-loop contribution of the 
$N-1$ $\eta$ fields, which is 
\be
\label{eta-1loop}
 {1\over 2} (N-1) \int{\rd^d k \over (2 \pi)^d} \log (k^2 + \chi). 
 \ee
The result in our renormalization scheme is given by
\be
\label{lop}
V_{(0)}(\phi, \chi)=    -{ \chi^2 \over 2 \kappa} +{\chi \phi^2 \over 2} +{ \mu^2 \chi \over \kappa} -{ \mu^4 \over 2 \kappa} +{1 \over 8 \pi} \chi \left(1-\log\left( {\chi \over m^2} \right) \right), 
\ee
a result first found in \cite{cjp}. The VEVs of $\phi, \chi$ at leading order in the $1/N$ expansion, which we denote by $\phi_{(0)}$, $\chi_{(0)}$, are obtained as the critical points of (\ref{lop}):
\be
\label{leading-min}
\ba
{\partial V_{(0)} \over \partial \phi }&= \phi \chi=0, \\
{\partial V_{(0)} \over \partial \chi}&= -{\chi \over \kappa} +{\phi^2 \over 2} +{ \mu^2 \over \kappa} -{1\over 8 \pi}\log\left( {\chi \over m^2} \right)=0. 
\ea
\ee
Solving for $\phi^2$ we obtain, 
\be
\label{leading-phi}
\phi_{(0)}^2={m^2 \over \kappa}+ { 2 \chi \over \kappa} +{1\over 4 \pi} \log\left( {\chi \over m^2} \right).  
\ee
The first equation in (\ref{leading-min}) has two solutions. The first one sets  $\phi_{(0)}=0$. This is the 
non-perturbative solution, which we will explore in the next section. 
The second one is 
\be
\label{chi0}
\chi_{(0)}=0, 
\ee
but this leads to a divergent value for $\phi_{(0)}^2$. In \cite{cjp} this is interpreted as an inconsistency coming 
from the assumption that there is a vacuum 
with spontaneously broken $O(N)$ symmetry. However, as shown in \cite{mr-new}, we can insist on 
following the perturbative approach of \cite{jevicki}. 
To do this, we pick the vacuum (\ref{chi0}) and we introduce an IR regulator $\epsilon$ as follows
\be
\label{leading-chi}
{\chi_{(0)} \over m^2}= \epsilon.  
\ee
We will take the limit $\epsilon \rightarrow 0$ at the end of the calculation, as in \cite{jevicki}.

The $1/N$ correction to the effective potential was first considered in \cite{root} and 
further studied in \cite{mr-new}. The main contribution comes from 
ring diagrams made out of the effective $\tilde \chi$ propagator. One finds
\be
\label{subleading-corr}
V_{(1)}(\phi, \chi)= {1\over 2} \int{\rd^d k \over (2 \pi)^d} \log\left[ -{1\over \kappa} \CS_{\tilde \chi}^{-1}\right]. 
\ee
 By using the effective potential at NLO in $1/N$ 
 it is also possible to calculate the $1/N$ corrections to the vevs of $\phi$, $\chi$ in the 
 perturbative vacuum. We will later need the correction to $\chi$, which reads
  \be
 \chi = \chi_{(0)}+ {1\over N} \chi_{(1)}+ \cdots, 
  \ee
where 
\be
\label{chi1}
\ba
\chi_{(1)}&= -\kappa  \int {\rd^d q \over (2 \pi)^d} {1\over (q^2+ \chi_{(0)}) (1+\kappa \Pi(q^2, \chi_{(0)}))+\kappa \phi^2_{(0)}} \\
&=-\kappa  \int {\rd^d q \over (2 \pi)^d} {1\over q^2+ m^2 + \gamma m^2 \log\left( {q^2 \over m^2} \right)} + \CO(\epsilon). 
\ea
\ee

The ground state energy at NLO in $1/N$ can be calculated from the 
effective potential. We now summarize the main result of \cite{mr-new}. It is convenient to 
introduce a normalized, dimensionless ground state energy  
\be
\mathfrak{E}= {E_{\rm p} \over m^2}, 
\ee
where the subscript ${\rm p}$ means that this is obtained as a 
perturbative series in the coupling constant. 
This quantity has as well a $1/N$ expansion of the form 
\be
\mathfrak{E}= N \mathfrak{E}_{(0)}+ \mathfrak{E}_{(1)}+ \cdots
\ee
Then, one finds \cite{mr-new} 
  \be
  \label{tot-final}
 8 \pi \mathfrak{E}_{(1)}(\gamma) = 1-{ \pi^2 \gamma \over 6}+  \int_0^\infty\left\{ \log\left[ 1+\gamma  { \log \left(t\right) \over t+1} \right] 
-\gamma { \log \left(t\right) \over t+1}  \right\} \rd t. 
\ee
This has to be understood as formal power series in $\gamma$, so we first 
expand the integrand in $\gamma$ and then we integrate order by order. The resulting series has the form 
\be
\label{e1-series}
8 \pi \mathfrak{E}_{(1)}(\gamma) =  1- {\pi^2 \over 6} \gamma + \sum_{n\ge 2}c_n \gamma^{n},  
 \ee
where the coefficients $c_n$ are given by 
 \be
 \label{cnco}
 c_n ={(-1)^{n-1}  \over  n} \int_0^\infty \left( {\log(t) \over 1+t} \right)^n \rd t, \qquad  n \ge 2.  
 \ee
 Explicitly, we find 
 \be
 \ba
8 \pi \mathfrak{E}_{(1)}(\gamma) &=  1- {\pi^2 \over6} \gamma- {\pi^2 \over6} \gamma^2 - {\pi^2 \over6} \gamma^3 -\frac{\pi^2  \left(30+7 \pi ^2\right) }{180}\gamma
   ^4-\frac{\pi^2  \left(60+77 \pi ^2\right) }{360}\gamma ^5 + \CO(\gamma^6).
   \ea
   \ee
It is easy to verify that this series is factorially divergent, i.e. $c_n \sim n!$. The source 
of this growth is the behaviour of the integrand at $t=0$. Since $t=k^2/m^2$, where $k^2$ is the squared momentum, 
this is an IR renormalon, as explained in \cite{mr-new}. 
To unravel the renormalon structure, one notes that the obstruction to make sense of the integral in (\ref{tot-final}) is the singularity of the integrand on the positive real axis. 
This singularity occurs at the solution of the equation 
\be
\label{vanishing}
t+1+ \gamma \log(t)=0,
\ee
which we will denote by $t_\star (\gamma)$. It is given explicitly by 
\be
\label{x-ts}
t_\star(\gamma)= \gamma W \left( {1  \over \gamma} \re^{-1/\gamma}\right)=\sum_{\ell=1}^\infty {\ell^{\ell-1} \over \ell!} (-1)^{\ell-1} \gamma^{1-\ell} \re^{-\ell/\gamma}, 
\ee
where $W$ is Lambert's function. We note that $t_\star (\gamma)>0$ for $\gamma>0$. 
We can then deform the integration contour slightly above or below the positive real axis, so as to avoid this singularity. Let us define 
\be
\label{contours}
\CC_{\pm} = \re^{\ri \theta_{\pm}} \IR_{\ge 0}, 
\ee
 where $\theta_\pm $ is a small positive or negative angle, respectively. The integrals along $\CC_\pm$ give the lateral Borel 
 resummations of the formal power series $\mathfrak{E}_{(1)} (\gamma)$:
\be
\label{lateralE}
s_\pm \left( 8 \pi\mathfrak{E}_{(1)} \right)(\gamma)= 1-{ \pi^2 \gamma \over 6}+  \int_{\CC_\pm} \left\{ \log\left[ 1+\gamma  { \log \left(t\right) \over t+1} \right] 
-\gamma { \log \left(t\right) \over t+1}  \right\} \rd t.  
\ee
One has the following 
explicit expression for the Stokes discontinuity as we cross the positive real axis:
\be
\label{discE}
(s_+ -s_-)\left(8 \pi  \mathfrak{E}_{(1)} \right)(\gamma)= 2 \pi \ri  t_\star (\gamma). 
\ee
This indicates that there are Borel singularities, due to IR renormalons, at all positive integers in 
the Borel plane of the coupling constant $\gamma$. We will see in the next section that these renormalons are indeed signals of 
nonperturbative corrections to the perturbative result (\ref{e1-series}). 

There is an alternative expression for the perturbative series which will be very useful 
when comparing to the non-perturbative large $N$ result. Let us consider a Borel transform of the series appearing 
in (\ref{e1-series})\footnote{The Borel transform considered in section 2.3 of \cite{mr-new} is different from this one.}, 
\be
 \sum_{n \ge 2} {8 \pi c_n \over (n-1)!} \zeta^{n-1}=- \int_0^\infty \left( \re^{- {\log(t) \zeta \over 1+t}}- 1+ {\log t \over 1+t} \right) {\rd t \over \zeta}. 
\ee
This integral representation is convergent for $|\zeta|<1$, i.e. before we reach the first singularity of the Borel transform. The original series can then be recovered by an inverse Borel transform, and is given by the formal integral,  
\be
-\int_0^\infty \re^{-\zeta/\gamma} \left( \re^{- {\log(t) \zeta \over 1+t}}- 1+ {\log t \over 1+t} \right) {\rd \zeta \over \zeta} \rd t. 
\ee
We can now change variables $\zeta= (1+t) \nu$, $u= t \nu/\gamma$, and integrate over $u$. As a result of these manipulations we find the expression 
\be
\label{ennu}
\ba
8 \pi \mathfrak{E}_{(1)}(\gamma)&=\gamma \int_0^\infty \left\{ \re^{- \nu \sigma_0} \nu^{\nu-1} \Gamma(-\nu) + 
\re^{-\nu/\gamma} \left( {\gamma_E + \log(\nu)- \log(\gamma) \over \nu} + {1\over \nu^2} \right) \right\} \rd \nu 
\\
& + 1-{\pi^2 \gamma \over 6}, 
\ea 
\ee
where we have denoted
\be
\label{s0}
\sigma_0= {1\over \gamma} + \log(\gamma). 
\ee
We note that the integrand in (\ref{ennu}) is regular at $\nu=0$ but it has simple 
poles at all positive integers $\nu=\ell \in \IZ_{>0}$, of the form 
\be
\label{poles-nu}
\gamma \re^{- \nu \sigma_0} \nu^{\nu-1} \Gamma(-\nu) =  {\ell^{\ell-1} \over \ell!} {(-1)^{\ell-1} \over \nu-\ell}\gamma^{1-\ell} \re^{-\ell/\gamma}+\cdots
 \ee
These are due of course to the singularities in the Borel transform. The lateral Borel resummations (\ref{lateralE}) can be 
obtained by deforming the integration contour in (\ref{ennu}). By using (\ref{poles-nu}), it is easy to see that the discontinuity across the positive real axis in the integral representation (\ref{ennu}) agrees with (\ref{discE}).

\subsection{The perturbative two-point function}

An interesting observable in this theory is the two-point function of the $\bPhi$ field. 
Due to the results of \cite{elitzur,david-cancel} on the non-linear sigma model, 
we expect that $O(N)$-invariant correlation functions will not have IR divergences. 
Let us then consider the connected, $O(N)$-invariant two-point function:
\be
\label{propa-N}
N S (p^2)=\int \rd^d x \, \re^{\ri p\cdot x} \left( \langle  \bPhi(x) \cdot \bPhi(0) \rangle - \langle  \bPhi(0) \rangle^2 \right), 
\ee
where we have extracted a factor $N$ so that $S(p^2) \sim 1$ at large $N$. Note that by considering the connected 
part, we remove the divergent contribution of the vev $\phi^2$. Since $\langle \xi \rangle=0$, 
we find
 \be
 \langle  \bPhi(x) \cdot \bPhi(0) \rangle - \langle  \bPhi(0) \rangle^2  =  \langle \xi(x) \xi(0) \rangle +\sum_{j=1}^{N-1} \langle \eta_j(x) \eta_j(0) \rangle. 
 \ee
 The perturbative self-energy at NLO in the $1/N$ expansion, $\Sigma^p_{(1)} (p^2; \gamma)$, is defined as 
 \be
 S^{-1}(p)= p^2+ {1\over N} \Sigma^p_{(1)} (p^2; \gamma)+ \CO(N^{-2}), 
 \ee
but it is more convenient to consider the normalized, dimensionless self-energy 
 \be
\label{dless-se}
 \mathfrak{S}(\mfp^2; \gamma)= {1 \over m^2}\Sigma^p_{(1)} (\mfp^2; \gamma), 
 \ee
 which is a function of $\gamma$ and the dimensionless quantity
\be
\mfp^2={p^2 \over m^2}. 
\ee
 We will now calculate $\mathfrak{S}(\mfp^2; \gamma)$ as a formal power series in $\gamma$, 
 at all loops. This is done with the IR regulator (\ref{leading-chi}), and we will take $\epsilon \rightarrow 0$ at the end of the calculation. As we will see, the contributions of the $\xi$ and the $\eta$ fields are separately IR divergent, but their sum is finite. 
 Let us first consider the two-point function of the $\eta$ fields, 
 \be
 S_{ij}(p^2)= \int \rd^d x \, \re^{\ri p\cdot x}  \langle \eta_i (x) \cdot \eta_j (0) \rangle. 
 \ee
  \begin{figure}[!ht]
\leavevmode
\begin{center}
\includegraphics[height=3.5cm]{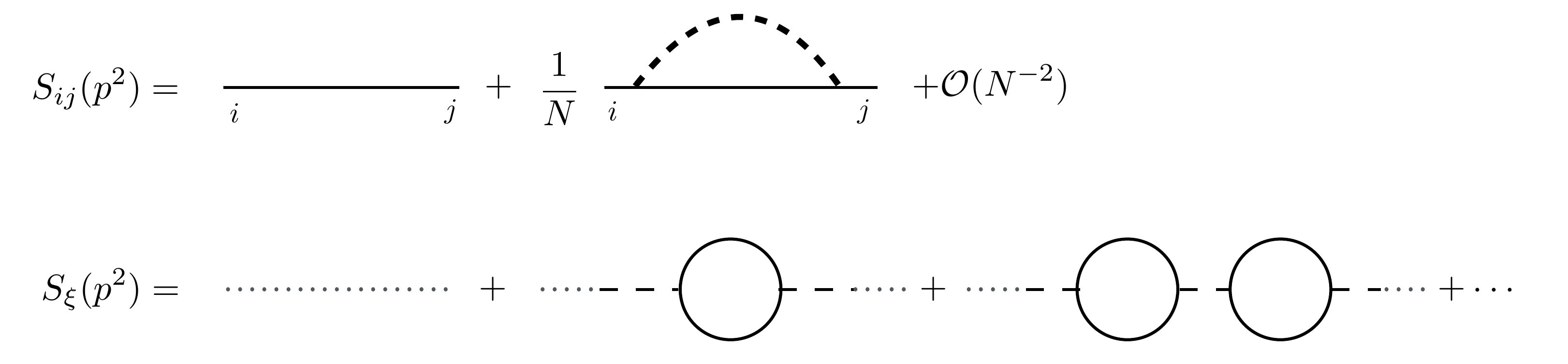}
\end{center}
\caption{The diagrams contributing to the propagator of the $\eta$ fields (in the first line) and to the propagator of the $\xi$ fields (in the 
second line). The thick dashed line in the first diagram is the effective propagator of the $\tilde \chi$ field, given by $S_{\tilde \chi}$. }
\label{two-point}
\end{figure} 
At next-to-leading order in $1/N$ it is given by the diagrams shown in the first line of 
\figref{two-point}. One important observation is that in the free propagator $1/(p^2+ \chi)$ we have to use the vev 
of $\chi$ at NLO in $1/N$. By using the explicit correction 
given in (\ref{chi1}) we find   
\be
\label{sij}
S_{ij}(p^2)= {\delta_{ij} \over p^2+ \chi_{(0)}}+ {1\over N} {\delta_{ij} \over \left(  p^2+\chi_{(0)} \right)^2} 
 \int {\rd^d q \over (2\pi)^d} \left(  {\CS_{\tilde \chi}(q^2)\over (p+q)^2+ \chi_{(0)}} - {\CS_{\tilde \chi}(q^2)\over q^2+ \chi_{(0)}} \right)+ \CO(N^{-2}). 
 \ee
 The subtraction inside the integral is conceptually important, 
 since it means that the $1/N$ correction to the self-energy of the 
$\eta$ particle {\it vanishes} at $p=0$, i.e. the $\eta$ particles are massless 
order by order in the perturbative $1/N$ expansion. In the two-dimensional case this result is only formal 
since, as we will see in a moment, $S_{ij}(p^2)$ is IR divergent, but it will eventually lead to the vanishing of the pole mass in the function $S(p^2)$. We obtain 
$S_\eta(p^2)$ by summing over the $N-1$ $\eta$ particles:
\be
\ba
S_\eta (p^2)&={1\over N}  \sum_{j=1}^{N-1} S_{ij}(p^2)= {1 \over p^2+ \chi_{(0)}}\\
& +{1\over N} \left\{ -{1\over p^2+ \chi_{(0)}}  +{1\over \left(  p^2+\chi_{(0)} \right)^2} 
 \int {\rd^d q \over (2\pi)^d} \left(  {\CS_{\tilde \chi}(q^2)\over (p+q)^2+ \chi_{(0)}} - {\CS_{\tilde \chi}(q^2)\over q^2+ \chi_{(0)}} \right) \right\}+ \CO(N^{-2}). 
 \ea
 \ee

Let us now calculate the contribution of the $\xi$ field. By summing all bubble diagrams in \figref{two-point} one finds
 \be
 \label{chi-cor}
N S_\xi (p^2)= {1\over p^2 + \chi_{(0)} + \kappa \phi_{(0)}^2}+ D_{\tilde \chi\xi}^2 (p) {\Pi(p^2, \chi) \over 1-\Pi (p^2, \chi) D_{\tilde \chi}(p^2)}, 
 \ee
 and the invariant $O(N)$ two-point function is given by the sum 
 \be
 S(p^2)= S_\eta (p^2)+ S_\xi(p^2). 
 \ee
Both $S_\eta(p^2)$ and $S_\xi (p^2)$ diverge separately as $\epsilon \to 0$, but their sum is finite. To see this, let us 
first consider $S_\xi (p^2)$, which is simpler.  The first term in the r.h.s. of (\ref{chi-cor}) will vanish in the limit $\epsilon \to 0$ since $\phi_{(0)}^2$ diverges logarithmically in this limit. The second term gives 
 \be
 \ba
& {\kappa^2 \phi_{(0)}^2  \over p^2 + \chi_{(0)} + \kappa \phi_{(0)}^2} {\Pi(p^2, \chi_{(0)})
 \over p^2+ m^2  + \gamma m^2  \log\left({p^2 \over m^2} \right)  }+\CO(\epsilon)\\
&=  {\kappa \Pi (p^2, \chi_{(0)}) \over p^2+ m^2  + \gamma m^2  \log\left({p^2 \over m^2} \right) }- {\kappa (p^2 + \chi_{(0)})\Pi(p^2, \chi_{(0)}) \over p^2 + \chi_{(0)}+ \kappa \phi_{(0)}^2} {1\over p^2+ m^2  + \gamma m^2  \log\left({p^2 \over m^2} \right) } + \CO(\epsilon). 
\ea
\label{xixi-int}
\ee
 We now use (\ref{propa-limit}). The first term in (\ref{xixi-int}) 
 is logarithmically divergent as $\epsilon \to 0$, while the second one gives a finite result since 
\be
\lim_{\epsilon \to 0} {(p^2 + \chi_{(0)})\Pi(p^2, \chi_{(0)}) \over p^2 + \chi_{(0)}+ \kappa \phi_{(0)}^2} =-{1\over \kappa}, 
\ee
We obtain in the end  
 \be
N S_\xi (p^2)= -{\gamma \over p^2}  {\log(\epsilon)  \over \mfp^2+1+ \gamma \log(\mfp^2)}+ 
{1 \over  \mfp^2 +1 + \gamma \log(\mfp^2)}\left( {1\over m^2}+  \gamma {\log(\mfp^2)\over p^2} \right) + \CO(\epsilon). 
\ee

Let us now consider $S_\eta (p^2)$. The only non-trivial contribution to this quantity is 
\be
\label{nt-ser}
\ba
&{1\over p^4}  \int {\rd^d q \over (2\pi)^d} {\CS_{\tilde \chi}(q^2)\over (p+q)^2+ \chi_{(0)}} \\
& \quad = - {\kappa \over p^4} 
 \sum_{n \ge 0} (-\gamma)^n  \int {\rd^d q \over (2\pi)^d} {q^2+ \chi_{(0)} \over (p+q)^2+ \chi_{(0)} } {m^{2n} \over (q^2+ m^2)^{n+1}} \log^n\left( {q^2 \over m^2} \right), 
 \ea
 \ee
where we have used (\ref{Schi-lim}). Let us note that the term with $n=0$ is 
UV divergent and has to be handled separately, by using the regularization (\ref{reg-int}). In terms of the 
dimensionless variable $y=q^2/m^2$, the series (\ref{nt-ser}) can be written as 
 \be
  \label{perts-etas}
 \ba
 &{1\over p^2} {\gamma \over \mfp^2} \left( \log(\epsilon)+ \int_0^\infty {\rd y \over {\sqrt{(y+ \mfp^2+ \epsilon)^2- 4 y \mfp^2}}}  {1 \over y+1}  \right)\\
 & + {1 \over p^2} {1\over \mfp^2} \sum_{n \ge 1} (-\gamma)^{n+1}\int_0^\infty {\rd y \over {\sqrt{(y+ \mfp^2+ \epsilon)^2- 4 y \mfp^2}}} {y\log^n(y) \over (y+1)^{n+1}},
 \ea
\ee
To shorten notation it is convenient to introduce the functions 
 \be
g_0(y)= {1\over y+1}, \qquad g_n(y)= {y \log^n(y) \over (y+1)^{n+1}}, \quad n \ge 1.
\ee
We will then write the $1/N$ correction to $S_\eta(p^2)$ as 
 \be
 {1\over p^2} {1 \over \mfp^2}\sum_{n\ge -1} (-\gamma)^{n+1} \CG_n(\mfp^2, \gamma; \epsilon), 
 \ee
 where 
 \be
 \label{In-int}
 \ba
  \CG_{-1}(\mfp^2; \epsilon)&= - \mfp^2, \\
  \CG_0 (\mfp^2; \epsilon)&=\log(\epsilon)+ \int_0^\infty {g_0(y) \over {\sqrt{(y+ \mfp^2+ \epsilon)^2- 4 y \mfp^2}}}  \rd y, \\
  \CG_n (\mfp^2; \epsilon)&=\int_0^\infty {g_n(y) \over {\sqrt{(y+ \mfp^2+ \epsilon)^2- 4 y \mfp^2}}}  \rd y-\int_0^\infty {g_n(y) \over y} \rd y , \qquad n \ge 1.  
  \ea
  \ee
The second term in the last line comes from the last term in (\ref{sij}) (the contribution of this term to $\CG_0$ vanishes due to our regularization). With the exception of $\CG_{-1}$, the quantities $\CG_n$ are IR divergent when $\epsilon \to 0$. The divergence in (\ref{perts-etas}) can be easily extracted, and one finds
\be
\label{div-fin}
\CG_n (\mfp^2; \epsilon)= -   \log(\epsilon)  {\mfp^2 \log^n(\mfp^2) \over (\mfp^2+1)^{n+1}} + \CJ_n (\mfp^2)+ \CO(\epsilon), 
\ee
for $n \ge 0$. In the $1/N$ correction to $S_\eta(p^2)$, these divergent parts add up to  
\be
- {1\over p^2}  \log(\epsilon) \sum_{n \ge 0} (-\gamma)^{n+1} {\log^n(\mfp^2) \over (\mfp^2+1)^{n+1}} = {\gamma \over p^2} {\log(\epsilon) 
\over \mfp^2 +1+ \gamma \log(\mfp^2)}, 
 \ee
which cancels the divergence of $S_\xi(p^2)$, as promised. The functions $\CJ_n(\mfp^2)$ 
can be obtained by using Mellin transforms, as 
explained in Appendix \ref{ap-finite}. A particularly useful expression is given in (\ref{J-ints}). The use of this expression gives a compact 
formula for the normalized self-energy (\ref{dless-se}), 
\be
\label{compact-sigma}
\ba
\mathfrak{S}(\mfp^2; \gamma)&=\mfp^2 { 1+\gamma \log(\mfp^2) \over \mfp^2+ 1+ \gamma \log(\mfp^2)}- \left( {2 \log(1+\mfp^2) \over 1+ \mfp^2} +  {\mfp^2 \log(\mfp^2) \over 1+ \mfp^2} \right) \gamma\\
&-\gamma  \int_0^\infty \left( {{\rm P} \over t- \mfp^2} +{1\over t} \right) \left(\frac{t}{t+1+\gamma  \log (t)}-\frac{t}{t+1} \right)  \rd t  \\
&- 2 \gamma  \int_{\mfp^2}^\infty \log\left( {t \over \mfp^2}-1 \right) 
\left(\frac{t}{t+1+\gamma  \log (t)}-\frac{t}{t+1} \right)' \rd t.
\ea
\ee
As in previous occasions, this expression has to be understood as a formal power series in $\gamma$, in which 
we first expand the integrand in $\gamma$ and then we integrate term by term. We find, at the very first orders, 
\be
\label{sigma-pers}
\ba
\mathfrak{S}(\mfp^2; \gamma)&= \sum_{n \ge 0} \mathfrak{S}_n(\mfp^2) \gamma^n\\
&= {\mfp^2 \over \mfp^2 + 1}- \left(  {2 \log(1+\mfp^2) \over 1+ \mfp^2} +  {\mfp^2 \log(\mfp^2) \over (1+ \mfp^2)^2 } \right) \gamma+ \CO(\gamma^2). 
\ea
\ee
We also have 
\be
 \mathfrak{S}_n(\mfp^2) = (-1)^n {\mfp^2 \log^n(\mfp^2) \over (\mfp^2 +1)^{n+1}} + (-1)^{n+1} \CJ_{n-1} (\mfp^2),  \qquad n \ge 2. 
\ee
By using the expression (\ref{compact-sigma}) it is easy to verify that 
\be
\lim_{\mfp^2 \to 0} \mathfrak{S}_n(\mfp^2)=0, \qquad n \ge 0.
\ee
This shows that the pole mass $m_p^2$ of the $\Phi$ field, which is defined as usual by $S^{-1}(p^2=- m_p^2)=0$, 
vanishes up to order $1/N$ in the large $N$ expansion, and at all orders in powers of the coupling constant $\gamma$. 
Although there are no Goldstone bosons in two dimensions, the $\Phi$ particles are massless in perturbation theory. 

We can use (\ref{compact-sigma}) to determine the nature of the perturbative expansion (\ref{sigma-pers}) 
for the self-energy. We first note that the obstruction to make sense of the integrals in (\ref{compact-sigma}) is the pole in the integrand 
at $t=t_\star (\gamma)$, as in the case of the ground state energy (\ref{tot-final}). We can then deform the integration contour slightly 
above or below the positive real axis, so as to avoid this pole. The resulting integrals are the lateral resummations of the formal power series $\mathfrak{S}(\mfp^2; \gamma)$:
\be
\label{Stokes}
\ba
s_\pm \left(\mathfrak{S}(\mfp^2) \right) (\gamma)&=\mfp^2 { 1+\gamma \log(\mfp^2) \over \mfp^2+ 1+ \gamma \log(\mfp^2)}- \left( {2 \log(1+\mfp^2) \over 1+ \mfp^2} +  {\mfp^2 \log(\mfp^2) \over 1+ \mfp^2} \right) \gamma\\
&-\gamma  \int_{\CC_\pm}  \left( {{\rm P} \over t- \mfp^2} +{1\over t} \right) \left(\frac{t}{t+1+\gamma  \log (t)}-\frac{t}{t+1} \right)  \rd t  \\
&- 2 \gamma  \int_{\mfp^2}^{\re^{\ri \theta_\pm } \infty} \log\left( {t \over \mfp^2}-1 \right) 
\left(\frac{t}{t+1+\gamma  \log (t)}-\frac{t}{t+1} \right)' \rd t. 
\ea
\ee
The Stokes discontinuity can be easily computed from (\ref{Stokes}). It is given by the trans-series
\be 
\left(s_+ -s_- \right)\left(\mathfrak{S}(\mfp^2) \right) (\gamma)= 2\pi \ri { \gamma \over \gamma + t_\star(\gamma)} \left( t_\star(\gamma)\pm  {t_\star^2(\gamma) \over t_\star(\gamma)- \mfp^2}\right), 
\ee
where the $\pm$ sign corresponds to the cases $\mfp^2>t_\star (\gamma)$ and $\mfp^2<t_\star (\gamma)$, respectively. It follows that there are Borel singularities at all positive integers, as in the case of the ground state energy.  The first exponential correction, which controls the leading contribution to the large order behavior of $\mathfrak{S}_n(\mfp^2)$, is simply $2 \pi \ri \, \re^{-1/\gamma}$, and independent of $\mfp^2$.  

The Borel singularities above are IR renormalons in the self-energy of the model, similar to the ones which are found in asymptotically free theories (see e.g. \cite{mm-trans} for a recent detailed analysis of the two-point function in the Gross--Neveu model). One difference however is that, in 
the asymptotically free case, the dependence on the momentum can be absorbed in the running coupling constant, and the power series for the two-point function has numerical coefficients. In this two-dimensional model, each coefficient $\mathfrak{S}_n(\mfp^2)$ is a complicated function of 
the momentum (see e.g. (\ref{J1})). This is sometimes called parametric resurgence, since the resurgent structure depends on a parameter, in this 
case the momentum $\mfp^2$. The parametric dependence in this case is very simple, and in particular 
the position of the singularities is momentum-independent.

\sectiono{Trans-series from the non-perturbative large $N$ solution}
\label{nonpert-sec}
\subsection{The non-perturbative large $N$ vacuum}

Let us now consider the other solution to the large $N$ vacuum equations, in which $\phi_{(0)}=0$. In \cite{cjp} this was  
regarded as the actual vacuum, in which no spontaneous symmetry breaking occurs, as 
required by the Coleman--Mermin--Wagner theorem. We will denote by 
\be
\chi_{(0)}= M^2
\ee
the vev of the auxiliary field in this vacuum, at leading order in $1/N$. From (\ref{leading-phi}) we find that it satisfies the equation
\be
\label{Meq}
M^2= -{m^2 \over 2} -{\kappa \over 8 \pi} \log \left( {M^2 \over m^2} \right).  
\ee
This is the analogue of the ``gap equation" obtained in the non-linear sigma model and the Gross--Neveu models. 
As shown in \cite{marco}, this equation can be solved as follows. Let us define the dimensionless quantity
\be
x= {M^2 \over m^2}. 
\ee
Then, (\ref{Meq}) gives
\be
\label{saddle-x}
x=-{1\over 2}-{\gamma \over 2} \log(x), 
\ee
which can be solved explicitly in terms of Lambert's function, 
\be
\ba
x&= {\gamma \over 2} W \left( {2 \over \gamma} \re^{-1/\gamma} \right) = \sum_{\ell =1}^\infty {(-2 \ell)^{\ell-1} \over \ell!} \re^{-\ell/\gamma}= \re^{-1/\gamma}-{2 \over \gamma} \re^{-2/\gamma}+{6 \over \gamma^2} \re^{-3/\gamma}+ \cdots
\ea
\ee
This is a purely non-perturbative quantity, which is in addition strictly positive for any $\gamma>0$ (in comparing with 
\cite{marco} we should note that here we are already making a choice of scheme, determined by (\ref{reg-int})). 
The vev of $\chi$ corresponds to a vev of $\bPhi^2$: it follows from (\ref{lag-chi}) that integrating out $\chi$ sets it 
equal to 
\be
\chi=\mu^2+ {\kappa \over 2N} \bPhi^2. 
\ee
We conclude that, at large $N$, 
\be
\label{true-vac}
\langle \bPhi \rangle=0, \qquad {1\over N} \langle \bPhi^2 \rangle= {1\over 4 \pi \gamma} \left(1+ 2 x \right),  
\ee
and we note that the first, perturbative 
term in the vev of $\bPhi^2$ agrees with the ``classical" vev (\ref{classm}). 
The ground state energy at leading order in the $1/N$ expansion is then obtained by 
evaluating (\ref{lop}) at this non-perturbative vacuum. After using (\ref{saddle-x}) we find, 
\be
\label{npE0}
{\cal E}_{(0)}= {E_{(0)} \over m^2}= -{1 \over 32 \pi} {1\over \gamma}+{x  \over 8 \pi} +{x^2 \over 8 \pi \gamma}, 
\ee
so already at leading order in the $1/N$ expansion we have non-perturbative corrections to the ground state energy.

Let us now consider the $1/N$ correction to this quantity. The propagator of the 
$\tilde \chi$ particle in the non-perturbative vacuum can be obtained 
by using standard techniques:
\be
\label{propa-tau}
\Delta^{-1}(k^2, M^2)= \kappa^{-1} +  \Pi (k^2, M^2), 
\ee
where $\Pi(k^2,M^2)$ is given in (\ref{Pif}). Then, the $1/N$ correction to (\ref{npE0}) 
is given by the one-loop contribution of the $\tilde \chi$ particle \cite{marco}:
\be
{\cal E}_{(1)}= {1\over 2 m^2} \int {\rd^d k \over (2 \pi)^d} \log(1+ \kappa \Pi (k^2, M^2)).
\ee
It is useful to introduce the notations of \cite{cr-dr} and write 
\be
\label{Feq}
1+ \kappa \Pi (k, M^2)= 1+ z F(y), \qquad F(y)=  {1\over y \xi} \log \left( {\xi +1 \over \xi-1} \right),
\ee
where
\be 
\label{ydef}
y={k^2 \over M^2}, \qquad \xi(y)= {\sqrt{1+ {4 \over y}}} 
\ee
and
\be
z(\gamma)={\gamma \over x(\gamma)}. 
\ee
We note that, in the perturbative regime $\gamma \ll 1$, $z \gg 1$ and is exponentially large. In terms of the variables in (\ref{Feq}) we have
\be
\label{E1y}
8 \pi {\cal E}_{(1)}(\gamma)= 
x \int_0^\infty  \log\left\{ 1+  {z\over y \xi} \log \left( {\xi +1 \over \xi-1} \right) \right\}\rd y. 
\ee

The integral in (\ref{E1y}) is clearly UV divergent. We will regularize it by using the sharp momentum cutoff (SM) scheme introduced in 
\cite{cr-dr, cr-review}, which is perhaps the most useful one for $1/N$ calculations in two-dimensional models. 
We first write
\be
\label{e1-finin}
\ba
8 \pi{\cal E}_{(1)}(\gamma)&=  x\int_0^\infty \left\{  \log\left(1+ {z\over y \xi} \log \left( {\xi +1 \over \xi-1} \right) \right)-   {z\over y \xi} \log \left( {\xi +1 \over \xi-1} \right) \right\} \rd y\\
& +  \gamma  \int_0^{\infty}  \log\left({\xi+ 1 \over \xi-1} \right){\rd y \over y \xi}.
\ea
\ee
It is easy to see that the first line is finite, so the UV divergence is in the term in the second line. The 
SM regularization goes as follows: we first introduce an UV 
cutoff in the integrals, $\Lambda$, and we expand the integrand around $y \to \infty$. We then 
subtract the part of the expansion responsible for the UV divergence. Since the 
subtracted part might lead to an IR divergence when we integrate it, we introduce as well an IR 
cutoff $\mu$ (not to be confused with the mass appearing in (\ref{spot})). The value of $\mu$ parametrizes 
the choice of scheme in the SM regularization, 
and we will choose it in such a way 
that we match the normal-ordering scheme used in perturbation theory. 
When applied to the second line in (\ref{e1-finin}) this method gives the regulated value
 \be
 \lim_{\Lambda \to \infty} \gamma \left\{  \int_0^{\Lambda^2/M^2} \log\left({\xi+ 1 \over \xi-1} \right){\rd y \over y \xi} - 
\int_{\mu^2/M^2}^{\Lambda^2/M^2} {\log(y) \over y} \rd y \right\}={\gamma \over 2} \log^2 \left( {M^2 \over \mu^2} \right).  
\ee
To match the scheme of perturbation theory, we consider the integral in the l.h.s. of (\ref{reg-int}). In the SM scheme this integral is regularized 
as follows
\be
\label{SM-loop}
\lim_{\Lambda \to \infty} {1\over 4 \pi} \left( \int_0^{\Lambda^2/M^2} {\rd y \over y+ a^2/M^2} - \int_{\mu^2/M^2}^{\Lambda^2/M^2} {\rd y \over y} \right)= 
-{1\over 4 \pi} \log\left( {a^2 \over \mu^2} \right),
\ee
which agrees with the r.h.s. of (\ref{reg-int}) precisely when $\mu=m$. We conclude that the renormalized vacuum energy is given by 
\be
\label{regE1}
8 \pi \CE^r_{(1)}(\gamma)=  x \int_0^\infty  \left\{  \log\left(1+  {z\over y \xi} \log \left( {\xi +1 \over \xi-1} \right) \right)-  {z\over y \xi} \log \left( {\xi +1 \over \xi-1} \right) \right\}\rd y+ {\gamma \over 2} \log^2(x). 
\ee
We can use the gap equation in the form (\ref{saddle-x}) to write the last term as a quadratic polynomial in $x$. Finally, the change of variable 
\be
\re^{\zeta}= {\xi + 1 \over \xi-1}
\ee
leads to the useful form 
\be
\label{ex-res}
\ba
8 \pi \CE^r_{(1)}(\gamma)&=  x \int_0^\infty  2 \sinh(\zeta)  \left\{  \log\left(1+ {z \zeta  \over 2 \sinh(\zeta)} \right)-  {z \zeta  \over 2 \sinh(\zeta)}\right\}\rd \zeta \\
&+ {1 \over 2 \gamma}+ {2 x \over \gamma} + {2 x^2 \over \gamma}.
\ea
\ee

\subsection{Trans-series for the ground state energy}
\label{subsec-ts}

\begin{figure}[!ht]
\leavevmode
\begin{center}
\includegraphics[height=4cm]{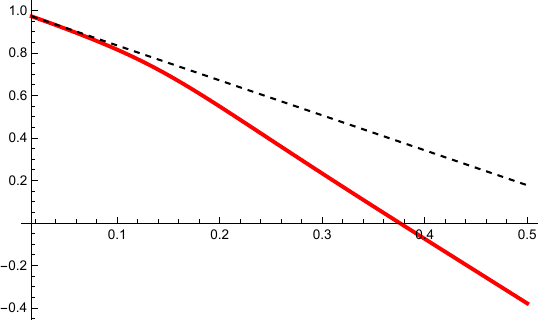}
\end{center}
\caption{The exact vacuum energy $8 \pi \CE^r_{(1)}(\gamma)$, as a function of $\gamma$ (red line), 
and its asymptotic behavior $1- \pi^2 \gamma/6$ as $\gamma \to 0$ (dashed black line).}
\label{exve-fig}
\end{figure}

Perhaps the most significant bonus of working at the non-perturbative large $N$ 
vacuum is that, order by order in $1/N$, one obtains actual 
{\it functions} of the coupling constant (after regularization), and not just 
formal power series. In particular, $\CE^r_{(1)}(\gamma)$, as defined by the integral expression (\ref{ex-res}), 
is a perfectly well-defined function of $\gamma$. It is plotted in \figref{exve-fig} as a function of $\gamma$, and it can be easily 
expanded in a series in $z$ around $z=0$, as noted in \cite{marco}, leading to a convergent series. In the theory 
with a positive squared mass, this series is an expansion in $g/m^2$, and it is a {\it bona fide} 
perturbative expansion. However, in the theory with a negative squared mass considered in this paper, 
$z$ is exponentially large. It can be easily checked that, in fact, 
the regime of small $z$ is never realized when $\gamma>0$, 
and in particular 
it is unrelated to the conventional perturbative regime. We then have to study the behavior 
of $\CE^r_{(1)}(\gamma)$ when $\gamma$ is very small and $z \gg 1$. 

Although the expression 
(\ref{ex-res}) was obtained around the non-perturbative large $N$ vacuum, which is very different from the 
perturbative vacuum considered in section \ref{pert-subsec}, one expects that the asymptotic expansion 
of the function (\ref{ex-res}) for small $\gamma$ is given by the perturbative series (\ref{e1-series}), i.e. 
\be
\label{asE1}
\CE^r_{(1)}(\gamma) \sim \mathfrak{E}_{(1)}(\gamma), \quad \gamma \to 0. 
\ee
A similar relationship can be found for the self-energies in e.g. the nonlinear sigma model and the Gross--Neveu models: 
the asymptotic expansion of the functions obtained in the non-perturbative vacua agree with the conventional perturbative expansion, 
up to a change of scheme. This was first 
argued in \cite{cr-dr} and later established with Mellin transform techniques in \cite{bbk} for the non-linear sigma model. 
It was recently verified in detail in \cite{mm-trans} in the case of the Gross--Neveu model. The method 
used in \cite{bbk} makes it possible to 
obtain as well the explicit trans-series of non-perturbative corrections, 
realizing explicitly in this way the renormalons as signals of non-perturbative physics 
in the perturbative series. 

The relation (\ref{asE1}) is not obvious from the formulae, but it can be verified numerically 
with some effort\footnote{The numerical calculation 
of the integral (\ref{ex-res}) is complicated for small values of $\gamma$.}. In \figref{exve-fig} we show 
for example that the linear approximation obtained by picking the first two terms in (\ref{e1-series}) matches 
correctly the exact result. Our goal now is to derive 
analytically the asymptotic result (\ref{asE1}), and to calculate as well the non-perturbative corrections. In other words, 
we want to obtain the full trans-series expression of the exact result (\ref{ex-res}). 
Our approach is therefore conceptually similar to the one in \cite{bbk,mm-trans}, although the trans-asymptotic 
analysis involves more elementary tools. As we will see, although (\ref{ex-res}) is deceptively simple, it leads to a 
trans-series including an infinite number 
of non-perturbative corrections, each of them being a non-trivial asymptotic series in the coupling $\gamma$.

Let us consider the function of $z$ given by 
\be
\CI(z)={1\over z} \int_0^\infty  2 \sinh(\zeta)  \left\{  \log\left(1+ {z \zeta  \over 2 \sinh(\zeta)} \right)-  {z \zeta  \over 2 \sinh(\zeta)}\right\}\rd \zeta, 
\ee
so that 
\be
8 \pi \CE^r_{(1)}=\gamma \CI(z(\gamma))+ {1 \over 2 \gamma}+ {2 x \over \gamma} + {2 x^2 \over \gamma}. 
\ee
We now use the elementary result 
\be
\int_0^\infty \left( \re^{-t (a+b)} \left(1+ a t \right) - \re^{-t b} \right) {\rd t \over t^2}= b \left( \log\left( 1+ {a \over b} \right)-{a \over b} \right)
\ee
to write
\be
\CI(z)= {1\over  z }  \int_0^\infty \rd \zeta \int_0^\infty \left( \re^{-t \zeta z - 2 t \sinh(\zeta)} \left(1+ z t \zeta \right) - \re^{- 2 t \sinh(\zeta)} \right) {\rd t \over t^2}. 
\ee
The integrals over $\zeta$ appearing here can be computed in closed form in terms of Bessel, Anger and Struve 
functions. We first have (\cite{GR}, 3.336, 1)
\be
\label{mast-int}
\int_0^\infty \re^{- \nu \zeta - \beta \sinh(\zeta)} \rd \zeta= {\pi \over \sin(\pi \nu)} \left( {\bf J}_\nu(\beta) - J_\nu (\beta) \right), 
\ee
where 
\be
  {\bf J}_\nu(\beta)={1\over \pi} \int_0^\pi \cos(\nu \theta- \beta \sin \theta) \rd \theta
  \ee
 is the Anger function, and $J_\nu(\beta)$ is the Bessel function of the first kind. The result (\ref{mast-int}) is valid when $\nu$ is not an integer. When $\nu=0$ we have (\cite{pk}, section 5.1.3)
  \be
  \int_0^\infty \re^{-  \beta \sinh(\zeta)} \rd \zeta= {\pi \over 2} \left( {\bf H}_0(\beta)- Y_0(\beta) \right), 
  \ee
  where ${\bf H}_0(\beta)$ is the Struve function and $Y_0(\beta)$ is the Bessel function of the second kind. 
  Let us define 
  \be
  \label{def-functions}
  \ba
  \CR(\nu,t)&= {\pi \over \sin(\pi \nu)} \left( {\bf J}_\nu(2t) - J_\nu (2t) \right), \\
  \CR(t)&= {\pi \over 2}\left( {\bf H}_0(2t)- Y_0(2t) \right). 
  \ea
  \ee
 We will need the series expansion around $t=0$ of $\CR(\nu,t)$ for generic $\nu$ \cite{pk}:
 \be
 \label{series-ex}
 \CR(\nu,t)= \sum_{k \ge 0} a_k(\nu)t^k + t^\nu \sum_{k\ge 0} b_k(\nu) t^{2k}, 
 \ee
where 
 \be
 \label{ab}
 a_k(\nu)= {(-1)^k\over 2} {\Gamma\left({\nu \over 2} -{k \over 2} \right) \over 
 \Gamma \left( 1+ {\nu \over 2} + {k \over 2} \right)} ={(-2)^k  \over 
 \prod_{i=0}^k (\nu+k -2i )}, \qquad b_k(\nu)={\Gamma\left( -\nu -k \right) \over k! }.
 \ee
 Using the above formulae, as well as the change of variables 
 \be
 t= {\nu \over z}
 \ee
 we obtain the representation
 \be
 \label{Iint}
 \CI(z)= \int_0^\infty \left\{ \CR\left(\nu, {\nu \over z} \right)- \nu \left( {\partial \CR(\nu, t) \over \partial \nu} \right)_{t=\nu/z} -\CR\left({\nu \over z} \right) \right\} {\rd \nu \over \nu^2}. 
 \ee
 We can write this integral as a series in $1/z$, by simply using the series expansion 
 (\ref{series-ex}). However, when we do this, each separate term in the series will be given by a singular integral. The singularities are due to the form of the functions $a_k(\nu)$, $b_k(\nu)$, and occur at integers on the positive real axis. We can however regulate these singular integrals by deforming the integration contour as in (\ref{contours}). We will then write: 
\be 
\label{z-ts}
\CI(z)= \sum_{n \ge 0} \CI^\pm_n (\sigma) z^{-n}, 
\ee
where we have introduced the variable 
\be
\sigma= \log(z). 
\ee
We have to distinguish two different cases. When $n \ge 3$, the integrals $\CI^\pm_n(\sigma)$ have the following structure 
\be
\CI^\pm_n (\sigma) =  \int_{\CC_\pm} \re^{-\sigma \nu} \CF_n(\nu) \rd \nu +  \int_{\CC_\pm} \CH_n(\nu)  
\rd \nu, 
\ee
where the functions $\CF_n (\nu)$, $\CH_n(\nu)$ can be obtained from the functions $a_k(\nu)$, $b_k(\nu)$. 
The functions $\CF_n(\nu)$ vanish when $n$ is odd, and when $n=2 k$, $k \ge 2$, they are given by 
\be
\CF_{2k}(\nu)= {1\over k!} \nu^{\nu+ 2k-2} \Gamma(-\nu-k) (\nu+ 2k), \qquad k \ge 2. 
\ee
The functions $\CH_n(\sigma)$ are given by
\be
\CH_n(\nu)= n a_n(\nu) \nu^{n-2}, \qquad n \ge 3,
\ee
In obtaining these results we have integrated by parts in (\ref{Iint}). 

The cases $n=0, 1,2$ are special, and caution has to be exercised in the integration by parts due to the 
singular behavior at $\nu=0$ of each separate  contribution to the integral (\ref{Iint}). It is therefore convenient to regulate the integrals by introducing an explicit lower cutoff $\epsilon>0$, and taking the limit $\epsilon \rightarrow 0$ at the end. We then find, 
\be
\label{eps-def}
\ba
 \CI^\pm _0(\sigma; \epsilon)&= \int_{\epsilon}^{ \re^{\ri \theta_\pm} \infty} \left\{ \re^{-\sigma \nu} \nu^{\nu-1} \Gamma(-\nu)+ {\gamma_E + \log(\nu)- \sigma \over \nu^2} \right\} \rd \nu+ {a_0(\epsilon) \over \epsilon}+ \re^{-\sigma \epsilon} \epsilon^{\epsilon-1} \Gamma(-\epsilon), \\ 
 \CI^\pm _1(\sigma; \epsilon)&= \int_{\epsilon}^{ \re^{\ri \theta_\pm} \infty} {a_1(\nu) \over \nu} \rd \nu + 2\left(-1+ \gamma_E + 2 \log(2) - \sigma \right) +2 \log(\epsilon), \\
\CI^\pm _2(\sigma; \epsilon)&= \int_{\epsilon}^{ \re^{\ri \theta_\pm} \infty}\left\{ \re^{-\sigma \nu} \nu^{\nu} (\nu+2)\Gamma(-\nu-1)+ 2 a_2(\nu) \right\} \rd \nu. 
\ea
\ee
We note that
\be
a_0(\nu)= {1\over \nu}, \qquad a_1(\nu)= {2 \over 1-\nu^2}, \qquad a_2(\nu)={4 \over \nu (\nu^2-4)}. 
\ee
To obtain the results in the first two lines of (\ref{eps-def}) we have to use (\ref{spec-int}) in Appendix \ref{app-ts}. 

We can now give explicit expressions for the $\epsilon \to 0$ limit of the above integrals. An elementary result that is useful to perform 
explicit calculations is that, if $f(\nu)$ has only simple poles at $\nu= \nu_n$, $n=1, \cdots, s$, on the real positive axis, then we have 
\be 
\label{rule-poles}
\int_{\CC^\pm} f(\nu) \, \rd \nu=  {\rm P} \int_0^\infty  f(\nu) \rd \nu
\mp \pi \ri \sum_{n=1}^s  {\rm Res}_{\nu= \nu_n} f(\nu). 
\ee
In the case $n=0$ one finds
\be
\label{CI0}
\CI^\pm _0(\sigma)=  \int_{\CC_\pm}  \re^{-\sigma \nu} \CF_0 (\nu) \rd \nu -{\sigma^2 \over 2} + \sigma- \sigma \log(\sigma) -{3 \log^2(\sigma) + \pi^2 \over 6}, 
\ee
 where 
 \be
 \CF_0(\nu)= \nu^{\nu-1} \Gamma(-\nu)+{1\over \nu^2} +{\gamma_E + \log(\nu) \over \nu}
 \ee
 is regular at $\nu=0$. In the case $n=1$, use of (\ref{rule-poles}) gives 
 \be
 \label{CI1}
\CI_1^\pm (\sigma)= -2 \sigma+ 2(\gamma_E + 2 \log(2)-1)\mp \pi \ri. 
\ee
 Finally, the integral with $\nu=2$ can be written as 
 \be
 \label{CI2}
  \CI^\pm _2(\sigma)=  \int_{\CC_\pm}  \re^{-\sigma \nu} \CF_2 (\nu) \rd \nu- 2 \log(\sigma)+   \int_{\CC_\pm}  \CH_2(\nu) \rd \nu, 
 \ee
 where 
 \be
 \CF_2(\nu)= \nu^{\nu} (\nu+2)\Gamma(-\nu-1)-{2\over \nu}, \qquad \CH_2(\nu)= {8  \over \nu(\nu^2-4)}+2 {\re^{- \nu} \over \nu}
 \ee
are manifestly regular at $\nu=0$.

 Let us now recapitulate our results so far. We have expressed the main contribution to 
 $\CE^r_{(1)}(\gamma)$ as an infinite series (\ref{z-ts}), in which each term $\CI_n^\pm (\sigma)$ 
 is given by an integral. Since $\sigma>0$, the integrands decay rapidly at infinity, and  
 $\CI_n^\pm (\sigma)$ are well defined functions of $\sigma$. We seem to have an ambiguity due to the 
 two choices of contour, and leading to two different functions $\CI^\pm_n(\sigma)$. These two 
 choices differ by the contribution of residues at the poles of the integrand, which is a purely 
 imaginary quantity. However, as we show in Appendix \ref{app-ts}, the difference cancels as we 
 sum over all $n\ge 0$ with the factor $z^{-n}$. This is exactly as in the example studied in \cite{bbk}, 
 and is a manifestation of the general mechanism of cancellation of renormalon ambiguities 
 discovered in \cite{david2}. 
 Finally, since $z^{-1} \sim \re^{-1/\gamma}$, the $\CI_n^\pm (\sigma)$ with $n\ge 1$ give non-perturbative 
 corrections to the perturbative result. 
 
 We want to extract now formal power series from the above integrals. Let us denote 
 \be
 h_n={\rm P} \int_0^\infty \CH_n(\nu) \rd \nu
 \ee
 and 
 \be
  h_{n,k}= {\rm res}_{\nu=k} \CH_n (\nu), \qquad n \ge 2, \quad  k=1, \cdots, n. 
 \ee
Then, by using (\ref{rule-poles}) we have
\be
\label{intH-pm}
\int_{\CC_{\pm}} \CH_n(\nu) \rd \nu= h_n\mp \pi \ri  \sum_{k=1}^n  h_{n,k}. 
\ee
We will extend this definition to $n=0,1$ by setting $h_0= 0$ and 
\be
h_1= 2(\gamma_E + 2 \log(2)-1), \qquad h_{1,1}=1. 
\ee
We consider now the asymptotic expansion of the following functions at large $\sigma$, 
\be
\CI_n^\pm (\sigma)\pm \pi \ri  \sum_{k=1}^n  h_{n,k} \sim \mathfrak{I}_n (\sigma), \qquad \sigma \to \infty. 
\ee
We note that the $\mathfrak{I}_n (\sigma)$ do not depend on the choice of integration contour, 
since the two choices differ in exponentially small terms $\re^{-k \sigma}$, $k \ge 1$. The 
$\mathfrak{I}_n (\sigma)$ are formal asymptotic series in $1/\sigma$, involving as well logarithmic 
terms of the form $\log^l(\sigma)/\sigma^k$. The explicit 
form of these series can be easily obtained from the expressions above, and one finds in this way the formal trans-series 
\be
\label{total-I}
 \mathfrak{I} (\sigma, z)  = \sum_{n \ge 0} \mathfrak{I}_n (\sigma)z^{-n}
\ee
However, in our original theory, the perturbative parameter is not $\sigma^{-1}$, but $\gamma$, and we have to rewrite (\ref{total-I}) in terms of it. One would expect that the resulting asymptotic expansions will have terms of the form $\gamma^k \log^l \gamma$, but this does not happen: they are formal power series in $\gamma$, involving at most a logarithmic term $\log(\gamma)$. We will now derive these explicit expansions. Let us note that  
\be
 \label{sig-ex}
 \sigma= \sigma_0+{2 x \over \gamma}, 
 \ee
where $\sigma_0$ was introduced in (\ref{s0}). To recover the conventional trans-series we have to do a formal expansion of $\mathfrak{I}_n(\sigma)$ around $\sigma =\sigma_0$.  

The first step is to verify that we reproduce the perturbative sector, i.e. that 
\be
\label{asym-proof}
8 \pi \mathfrak{E}_1 (\gamma)= \gamma \mathfrak{I}_0(\sigma_0) +{1\over 2 \gamma}. 
\ee
This can be easily checked analytically by using the expression 
(\ref{ennu}) and the calculation of (\ref{spec-int2}) in the Appendix \ref{app-ts}. 
This gives a proof of the asymptotic statement (\ref{asE1}). Note that the ambiguities due to 
renormalon singularities that appear in the resummation of the perturbative series, will 
cancel against some of the explicit ambiguities appearing in (\ref{intH-pm}). 

Let us consider the trans-series corrections. 
The functions $\mathfrak{I}_{n}(\sigma_0)$ for $n$ odd are trivial. We now show that the formal 
series $\mathfrak{I}_{2k}(\sigma_0)$, and their derivatives w.r.t. $\sigma_0$, can be expressed as 
formal power series in $\gamma$ (except for $k=1$, where we have a $\log (\gamma)$ term). To do this we just 
reverse the argument that led to (\ref{ennu}). We find, for the first two derivatives of $\mathfrak{I}_0(\sigma_0)$, 
\be
\ba
\mathfrak{I}'_0(\sigma_0)
&={1\over \gamma} \int_0^\infty \re^{-\zeta/\gamma} \rd \zeta \int_0^\infty  \left[ \re^{- {\log(t) \over 1+t}  \zeta } - 1  \right]{ \rd t \over 1+t} -{1\over \gamma} ,\\
\mathfrak{I}''_0(\sigma_0)
&=-{1\over \gamma} \int_0^\infty \re^{-\zeta/\gamma} \rd \zeta \int_0^\infty \zeta \re^{- {\log(t) \over 1+t}  \zeta }  {\rd t \over (1+t)^2} -1,
\ea 
\ee
Here and in subsequent formulae, the integral over $\zeta$ has to be understood as a formal Laplace transform.
For $\mathfrak{I}_2(\sigma_0)$ we find
\be
\ba
\mathfrak{I}_2(\sigma_0)
&={1\over \gamma} \int_0^\infty \re^{-\zeta/\gamma} \rd \zeta \int_0^\infty  \left[  {\zeta+ 2 (1+t) \over \zeta+1 +t}\re^{- {\log(t) \over 1+t}  \zeta } - 2  \right]{ \rd t \over 1+t}\\
& +2 \log(\gamma) -2 (\gamma_E + \log(2)). 
\ea 
\ee
For the remaining cases we have the single formula, 
\be
\label{genJ}
 \mathfrak{I}^{(p)}_{2k} (\sigma_0)={(-1)^p \over \gamma}
  \int_0^\infty \re^{-\zeta/\gamma} \rd \zeta \int_0^\infty  {\zeta^{2k-2+p } (\zeta+ 2k (1+t)) \over \prod_{i=1}^k (\zeta+ i (1+t))}\re^{- {\log(t) \over 1+t}  \zeta } { \rd t \over (1+t)^{k+p}} + h_{2k} \delta_{p0}, 
\ee
which is valid for $2k+p\ge 3$. Note that these formulae give the formal series in $\gamma$ 
through their explicit Borel transforms. These are obtained as integrals over $t$ which are convergent when $|\zeta|<1$, and 
can be easily computed as formal series in $\zeta$ around $\zeta=0$. The lateral resummations of the resulting  
series in $\gamma$ reconstruct the functions $\CI_n^\pm (\sigma_0)$: 
\be
\CI_{n}^\pm (\sigma_0)= s_\pm \left(  \mathfrak{I}_{n} \right)(\gamma) \mp  \pi \ri \sum_{k =1}^n h_{n,k}. 
\ee

With the above results we can promote the asymptotic expansion (\ref{asE1}) to a full trans-series asymptotics for the NLO ground state energy. 
We will write it as 
 \be
\CE_{(1)}^r(\gamma) \sim  \sum_{k \ge 0}  \mathfrak{E}_{(1)}^{(k)}(\gamma) x^k, 
 \ee
 where $\mathfrak{E}_{(1)}^{(0)}(\gamma)= \mathfrak{E}_{(1)}(\gamma)$ is the perturbative series (\ref{e1-series}). 
 We have chosen $x$ as the non-perturbative parameter, since it is the natural one coming from the gap equation, 
 but one could also use $\re^{-1/\gamma}$. The coefficients of $\mathfrak{E}_{(1)}^{(k)}(\gamma)$ as series in $\gamma$ 
 will be given by integrals akin to (\ref{cnco}). We find, for the first non-perturbative correction,
 \be
 \label{fts}
8 \pi \mathfrak{E}_{(1)}^{(1)}(\gamma)= \mathfrak{I}_1 (\sigma_0)+ 2   \mathfrak{I}'_0(\sigma_0) + {2 \over \gamma} \mp \pi \ri,  
 \ee
 where the third term in the r.h.s. comes from the second term in the last line of (\ref{ex-res}). Explicitly, 
 \be
  \label{fts-ex}
 \ba
 & 4 \pi \mathfrak{E}_{(1)}^{(1)}(\gamma)=-{1\over \gamma} - \log(\gamma)-1+ \gamma_E + 2 \log(2) + \sum_{n \ge 1}
 (-\gamma)^n \int_0^\infty {\log^n(t) \over (1+t)^{n+1}} \rd t \mp {\pi\ri  \over 2} \\
&= -{1 \over \gamma} -  \log(\gamma) -1+ \gamma_E+ 2 \log(2) +  {\pi^2 \over 6}  \gamma^2 + { \pi^2  \over 2} \gamma^3+ \left(\pi ^2+\frac{7 \pi ^4}{60}\right) \gamma^4+ \cdots   \mp {\pi \ri \over 2}.
\ea
\ee
This result can be verified numerically by considering the asymptotic behavior of the function
\be
\CE^r_{(1)}(\gamma)- s_\pm \left( \mathfrak{E}_{(1)} \right)(\gamma).  
\ee
As usual in these representations, the choice of sign for the last term in (\ref{fts}), $\mp \pi \ri/2$, must be correlated 
to the choice of lateral Borel resummation for the perturbative series, so that the ambiguity associated to the first 
renormalon cancels. 

As a final illustration, let us consider the second trans-series. It is given by  
\be
8 \pi \gamma^2 \mathfrak{E}_{(1)}^{(2)} (\gamma)= \mathfrak{I}_2(\sigma_0) +2 \mathfrak{I}'_1(\sigma_0)+ 2 \mathfrak{I}_0''(\sigma_0) +2 \gamma \mp \pi \ri. 
\ee
Explicitly, one finds
\be
\ba
8 \pi \gamma^2  \mathfrak{E}_{(1)}^{(2)} (\gamma)&=  -2 (\log(2)+ \gamma_E+3) + 2 \log (\gamma)+ \gamma+ {2 \pi^2 \gamma^2 \over 3}+ {5 \pi^2 \gamma^3 \over 3}\\
&  +  {\pi^2 \over 30} (95+ 14 \pi^2) \gamma^4+\CO(\gamma^5) \pm \pi \ri. 
   \ea
   \ee

 \subsection{The non-perturbative two-point function}

\begin{figure}
\centering
\includegraphics{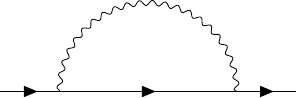}
\qquad \qquad
\includegraphics{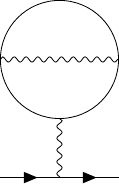}
\caption{Self-energy diagrams at large $N$. The wavy line corresponds to the large $N$ propagator of the $\tilde \chi$ field. }
\label{exact-selfenergy-diagrams}
\end{figure}%

Another quantity that can be easily computed in the non-perturbative large $N$ vacuum at NLO in $1/N$ is the self-energy of the $\Phi$ field. Its calculation is very similar to what is done in the non-linear sigma model and the Gross--Neveu models in \cite{cr-dr}, see \cite{marco,liu-ope} for further details. One finds
\be
\label{sigmaNP}
\Sigma_{(1)}(p)= \int {\rd^d k \over (2 \pi)^d} {\Delta (k^2,M^2) \over (p+k)^2+ M^2} - {\Delta(0,M^2)  \over 2}  \int {\rd^d k \over (2 \pi)^d} 
{\rd^d q \over (2 \pi)^d}{ 1 \over (q^2+ M^2)^2} {\Delta(k^2,M^2) \over (q+k)^2+ M^2}. 
\ee
This is non-perturbative and 
will produce an actual function of $\gamma$, after renormalization. We recall that $\Delta(k^2,M^2)$ is given by (\ref{propa-tau}). 
The first and second terms in the r.h.s. of (\ref{sigmaNP}) correspond to the two diagrams shown in \figref{exact-selfenergy-diagrams}. We can simplify the second term by noting that,
\be
\ba
{\partial \Delta^{-1}(k^2, M^2) \over \partial M^2}&=- \int {\rd^d q \over (2 \pi)^d}{ 1 \over (q^2+ M^2)^2} {\Pi(k^2, M^2) \over (q+k)^2+ M^2}\\
&=-{2 \over k^2 \xi^2} \left( \Pi (k^2, M^2) +{1\over 8 \pi M^2} \right),
\ea
\ee
and one finds the useful expression
\be
\ba
\Sigma_{(1)}(p)&= \int {\rd^d k \over (2 \pi)^d} {\kappa \over 1+ \kappa \Pi (k^2, M^2)} {1 \over (p+k)^2+ M^2} \\
& - \Delta(0,M^2)  \int {\rd^d k \over (2 \pi)^d} 
{\kappa \over 1+ \kappa \Pi (k^2, M^2)}  {1 \over k^2 \xi^2} \left( \Pi (k^2, M^2) +{1\over 8 \pi M^2} \right)
\ea
\ee
We also note that, in this case, 
\be
\Delta(0,M^2)=\left({1\over \kappa}+ {1\over 8 \pi M^2}\right)^{-1}= {2 \kappa \over  2+z}. 
\ee
After some massaging one finds
\be
\label{inf-sigma}
\ba
\Sigma_{(1)}(p)&= {\kappa \over 4 \pi} \int_0^\infty \left[ {1 \over 1+ z F(y)} \left(  {1\over {\sqrt{(1+ y+ v)^2-4 v y}}} 
-{z-2 \over z+2} {1 \over y+4} \right) -{4 \over z+2} {1\over y+4} \right] \rd y\\
& +{\kappa \over 2  \pi} {1\over 2+z}  \int_0^\infty {\rd y \over y+4},  
\ea
\ee
where $y$ is the variable introduced in (\ref{ydef}) and 
\be
v= {p^2 \over M^2} ={ \mfp^2 \over x}. 
\ee
The first line of (\ref{inf-sigma}) is finite, but the integral in the second line is UV divergent. We renormalize 
it by using again SM regularization. Its finite part is obtained from (\ref{SM-loop}) or directly from (\ref{reg-int}).

After all these calculations, we find a renormalized self-energy which can be written in terms of 
dimensionless quantities as
\be
\label{selfNP}
{\Sigma^{ r}_{(1)}(p)\over m^2} = S(\mfp^2;\gamma)+ \mathsf{m}, 
\ee
where
\be
\label{spg}
S(\mfp^2;\gamma)=  \gamma  \int_0^\infty {\rd y \over 1+ z F(y)} \left(  {1\over {\sqrt{(1+ y+ v)^2-4 v y}}} 
-{1\over {\sqrt{y^2+ 4 y}}} \right), 
\ee
and
\be
\mathsf{m}=  \gamma  \int_0^\infty \left[{ 1\over 1+ z F(y)} \left( {1\over {\sqrt{y^2+ 4 y}}}-{z-2 \over z+2} {1 \over y+4} \right) -{4 \over z+2} {1\over y+4} \right] \rd y - \gamma { 2 \log(4 x) \over 2+z}.
\ee
The function (\ref{spg}) satisfies the property 
\be
\label{smx}
S(-x;\gamma)=0,
\ee
where the evaluation point $\mfp^2=-x$ corresponds to $p^2=-M^2$ or $v=-1$. 
The non-perturbative two-point function of the $\Phi$ field at NLO in $1/N$ is then given by 
\be
{1\over m^2} S^{-1}_{\rm NP}(p^2)=\mfp^2+ x + {1\over N}(S(\mfp^2;\gamma)+ \mathsf{m}) + \CO(N^{-2}). 
\ee
Due to the property (\ref{smx}), the pole mass $m^2_p$ is given by 
\be
{m^2_p \over m^2}= x + {1\over N} \mathsf{m}+ \CO(N^{-2}), 
\ee
i.e. $\mathsf{m}$ gives the $1/N$ correction to the pole mass. A similar calculation of the pole mass 
for the non-linear sigma model and the Gross--Neveu model was done in \cite{cr-dr}. We have seen that the pole mass 
vanishes in perturbation theory, at this order in $1/N$, so $\mathsf{m}$ must be purely non-perturbative. In addition, the asymptotic expansion of  the exact function $S(\mfp^2;\gamma)$, at small $\gamma$, must be given by the formal power series (\ref{sigma-pers}), i.e. 
\be
\label{SmfS}
S(\mfp^2;\gamma) \sim \mathfrak{S}(\mfp^2; \gamma). 
\ee
We have checked this asymptotic behavior numerically. Using the results in \cite{liu-ope} for the theory 
with a positive squared mass, it is possible to verify (\ref{SmfS}) \cite{liu-pc}. 

In contrast to what happens in asymptotically free theories, where the $1/N$ correction to the pole mass is proportional to the dynamically generated scale \cite{biscari,cr-dr,fnw2}, here $\mathsf{m}$ is given by a (resummed) trans-series and in particular it has a non-trivial dependence on the coupling constant $\gamma$. Let us give some details on this, although the derivation is similar to the one of the ground state energy. If we change variables as we did in (\ref{ex-res}), we find
\be
\ba
\mathsf{m}&= 2  \gamma \int_0^\infty \left\{ {1 \over z \zeta + 2 \sinh(\zeta)} \left[\sinh (\zeta)- {z-2 \over z+2} (\cosh(\zeta)-1) \right]-{2 \over z+2} \tanh(\zeta/2) \right\} \rd \zeta \\
&-\gamma { 2 \log(4 x ) \over 2+z}. 
\ea
\ee
We now use the integration trick
\be
{1\over a}= \int_0^\infty  \re^{-a t}\, \rd t
\ee
to write 
\be
\ba
\mathsf{m}&=   \gamma \int_0^\infty \left[ \left(1-{2 \over z} \right) \CR\left( \nu, {\nu \over z} \right)-{1\over z} \left( {\partial \CR (\nu, t) \over \partial t} \right)_{t= \nu/z} -{z-2 \over z+2} {1\over \nu} -{4 \over z+2} { \re^{-4 \nu/z} \over \nu} \right] \rd \nu\\
&+ m(\gamma, z)
\ea
\ee
where
\be
m(\gamma, z)=-\gamma { 4 \log(2 ) \over z+2}+{2 \over z}{2 \gamma +z \over z+2}=
\sum_{n \ge 1} m_n(\gamma) z^{-n}. 
\ee
As in the case of the ground state energy, we can calculate the trans-series expansion of this quantity by doing an expansion of the integrand in $1/z$. One obtains in this way 
\be
\mathsf{m}= \sum_{n \ge 1} \mathsf{m}^\pm_n z^{-n}, 
\ee
where 
\be
 \mathsf{m}^\pm_n=\gamma  \int_{\CC_\pm} \left( \re^{-\sigma \nu} \CM_n(\nu)+ \CN_n(\nu) \right) \rd \nu -2 \delta_{n1} \log(\sigma)+  2^{n+1} (-1)^n  \gamma \sigma+ m_n(\gamma).
 \ee
  The functions $\CM_n(\nu)$, $ \CN_n(\nu)$ are regular at $\nu=0$ but have poles at non-negative integers. They are given as follows. For $n=1$ we have
  \be
  \ba
  \CM_1(\nu)&=-2\left( \nu^\nu \Gamma(-\nu)+{1\over \nu} \right),  \qquad \CN_1(\nu)&={2 \over \nu}\left(1-\re^{- \nu}\right)-{2 \over1+ \nu}. 
  \ea
  \ee
For $n \ge 2$, we have
\be
\ba
\CM_n(\nu)&=  -\nu^{\nu+n-1} c_n b_{[{n+1 \over 2}]}(\nu),\\
\CN_n(\nu)&= \nu^{n-1} \left( (\nu-n) a_n(\nu)- 2 a_{n-1}(\nu)\right) -2^{n+1} (-1)^n {1-\re^{- \nu} \over \nu},
\ea
\ee
where $c_n=2$ if $n$ is odd, and $c_n=n$ if $n$ is even. We recall that $a_k(\nu)$, $b_k(\nu)$ were introduced in 
(\ref{ab}). With the above results we can write a trans-series expansion for $\mathsf{m}$, of the form 
\be
\mathsf{m}\sim \sum_{n \ge 1} \mathfrak{M}_n(\gamma) x^n, 
\ee
and the formal power series $\mathfrak{M}_n(\gamma)$ can be calculated by using the same techniques that we used 
in the case of the ground state energy. One finds for example, for the first correction, 
\be
\mathfrak{M}_1(\gamma) = -{2 \over \gamma}- 2 \log(\gamma) + 2(\gamma_E + 2 \log(2)) + 
2 \sum_{n \ge 1} (-\gamma)^n \int_0^\infty {\log^n(t) \over (1+t)^n} \rd t, 
\ee
which is almost identical to the first correction to the ground state energy in (\ref{fts-ex}). 

\sectiono{Conclusions and open problems}

In this paper we have studied the renormalon appearing in the ground state energy of the quartic scalar theory with negative mass squared. 
This renormalon was discovered in \cite{mr-new} from a purely perturbative approach, and its connection to the 
non-perturbative result obtained e.g. in \cite{marco} in the large $N$ 
expansion was not completely clear. Similar results for the self-energies of solvable two-dimensional models \cite{cr-dr,bbk,mm-trans} suggested that the 
perturbative result of \cite{mr-new} should give the asymptotic expansion of the exact ground state energy obtained in the non-perturbative vacuum. 
We have verified this in detail and found in addition the complete series of trans-series corrections to the non-perturbative result. This shows that the 
IR renormalons obtained in \cite{mr-new} are the smoking gun for this infinite series of non-perturbative corrections. An additional lesson of this exercise is that 
the extraction of the asymptotic perturbative series and its corrections from a non-perturbative result might be far from obvious. In that sense, the perturbative 
expansion around the ``false" vacuum discussed in \cite{jevicki,mr-new} is a powerful method to obtain asymptotic results. 

In addition, we have studied the $O(N)$-invariant two-point function of the $\Phi$ field at the perturbative and the non-perturbative level. We have 
extended Jevicki's approach in \cite{jevicki} and verified IR finiteness once all diagrams are added. As in asymptotically 
free theories, the self-energy has IR renormalon singularities at NLO in $1/N$ which can be analyzed in detail. The corresponding pole mass 
vanishes at all orders in perturbation theory but it is given by a non-perturbative trans-series which can be explicitly computed. Although we have not 
worked out the non-perturbative corrections to the full two-point function, which become more complicated, 
the techniques introduced recently in \cite{liu-ope} could be used to 
find them explicitly. This is yet another example where the perturbative calculation, although intricate due to the presence IR divergences, gives an easier 
access to the asymptotic series than the non-perturbative result. 

Our result makes it clear that the physical origin of all these non-perturbative corrections is the 
structure of the true vacuum, 
summarized in (\ref{true-vac}): 
in this vacuum, the condensate associated to $\Phi$ vanishes (in contrast to what is assumed in 
perturbation theory), while the condensate of the 
operator $\Phi^2$ develops a non-perturbative correction to its classical value. Perhaps the 
most interesting problem open by this investigation would be to reproduce the 
non-perturbative corrections directly from an OPE/condensate calculation, as it was done 
recently in \cite{mm-trans} for the self-energy of the Gross--Neveu model. Similar calculations 
have been done in the past in the somewhat 
inverse situation, in which one wants to reproduce 
results in the spontaneously broken theory by using the OPE with condensates in 
the disordered phase \cite{scalar-ope,svz-ope}. Here, we would like 
on the contrary to go back to the disordered phase, and to turn on in addition a 
non-perturbative condensate for the $\Phi^2$ operator. It is likely that the recent 
results of \cite{liu-ope} will be very helpful for such a calculation, at least in the case 
of the two-point function, in which we can use the OPE. In the case of the  
ground state energy we cannot use the OPE directly, and we would need a 
generalization of the ideas and techniques of \cite{svz} to reproduce the trans-series 
obtained in this paper. This remains an important and challenging problem. 
In this respect, it is interesting to note that, in the first non-perturbative correction (\ref{fts-ex}), the infinite 
sum in the first line is 
precisely what would one obtain by inserting an appropriate ``condensate" in the ring diagrams 
of the perturbative series. 
Thanks to formulae like (\ref{genJ}), it might be possible to decode as well the higher corrections in terms of 
diagrams with condensates. 

Our analysis shows that, although the renormalons in the perturbative series of the ground state energy 
are ``smoking guns" for non-perturbative corrections, they do not capture them in any detail. As already 
shown in \cite{mr-new}, the singularities of the Borel transform of the perturbative series are poles, 
and their associated trans-series are trivial. In contrast, the non-perturbative corrections involve non-trivial series. 
For example, at the first non-perturbative order $x \approx \re^{-1/\gamma}$, the non-perturbative correction is given by 
(\ref{fts-ex}), while the renormalon only captures the ambiguous term $\mp \pi \ri/2$. This situation is typical of the 
large $N$ analysis of \cite{bbk,mm-trans,dmss}, and also of the standard renormalon analysis at large number of 
flavours $N_f$ in QCD. It is another instance of the ``weak" resurgence scenario in \cite{dmss}. However, as pointed out 
in \cite{dmss}, this might be an artifact of the large $N$ expansion, and perhaps at finite $N$ we 
have a ``strong" resurgence scenario, i.e. it might happen that at finite $N$ the resurgent 
structure of the perturbative series encodes all the needed corrections. This has been recently shown to be the 
case for the perturbative series of the ground state energy in the Lieb--Liniger model \cite{bbal}, which is a non-relativistic 
version of the $O(2)$ quartic model studied here, as noted in \cite{jevicki,mr-new}. 

Finally, there have been proposals for a semiclassical understanding of renormalon corrections, 
see e.g. \cite{matt} for a recent one. In our view, this and previous proposals remain at the very least vague, 
and they have not made any contact with the -by now extensive- collection of explicit renormalon trans-series 
obtained both at large $N$ and finite $N$ in many two-dimensional models 
(see e.g. \cite{tomas-thesis,serone-review} for reviews in the integrable case). The example in this paper (together with 
the self-energy of the Gross--Neveu model studied in \cite{mm-trans}) is perhaps the simplest case where a 
proposal for a semiclassical theory of renormalons could prove its worth.

\section*{Acknowledgements}
We would like to thank Yizhuang Liu, Ramon Miravitllas, Tom\'as Reis and Marco Serone for sharing 
their insights on this subject. Thanks also to Tom\'as Reis and Marco Serone for their comments 
on a preliminary version of this paper, and Yizhuang Liu for communicating his results on the two-point function. 
This work has been supported in part by the ERC-SyG project 
``Recursive and Exact New Quantum Theory" (ReNewQuantum), which 
received funding from the European Research Council (ERC) under the European 
Union's Horizon 2020 research and innovation program, 
grant agreement No. 810573.

\appendix

\sectiono{Some computational details}
\label{ap-finite}

Since some of the calculations in this paper are relatively involved, and not completely obvious 
(at least to the author), we have collected in this Appendix some details which might be useful to the reader. 
\subsection{Perturbative calculation of the two-point function}

We want to calculate the integrals appearing in the expression (\ref{In-int}) for $\CG_n(\mfp^2; \epsilon)$, 
as $\epsilon \to 0$. To do this, we adapt a trick from \cite{bbk}. We recall that the Mellin transform of a function $f(y)$ is defined as 
\be
\CM[f(y);s]=\int_0^\infty y^{s-1} f(y) \rd y.
\ee
In \cite{mellin}, 2.64, we find
\be
\int_0^\infty  \frac{y^{s}}{\sqrt{(x+y+\epsilon)^2-4xy}}\rd y= (x+\epsilon)^{s} B(s+1, -s) {}_2 F_1 \left( s+1, -s, 1; {x\over x+\epsilon} \right). 
\ee
We also have the expansion
\be
\ba
&(x+\epsilon)^{s}  {}_2 F_1 \left( s+1, -s, 1; {x\over x+\epsilon} \right)=\\
-& {x^s \over \Gamma(-s)} \sum_{k \ge 0} {(-s)_k \over k!^2 \Gamma(s+1-k)} \left\{ \log(\epsilon) - \log(x) -2 \psi(k+1)+ \psi(-s+k) + \psi(s+1-k) \right\} \left( {\epsilon \over x} \right)^k. 
\ea
\ee
In the limit $\epsilon \to 0$ we can just keep the term $k=0$ in the above series. We then get, 
\be
\int_0^\infty \rd y \frac{y^{s}}{\sqrt{(x+y+\epsilon)^2-4xy}} = -x^s  \left\{ \log(\epsilon) - \log(x) +2\gamma_E+ \psi(-s) + \psi(s+1) \right\}+\CO(\epsilon). 
\ee
Therefore, we can calculate the integrals appearing in (\ref{perts-etas}), in the limit $\epsilon\rightarrow 0$, by an inverse 
Mellin transform: 
\be
\ba
&\int_0^\infty {g_n(y) \over {\sqrt{(y+ \mfp^2+ \epsilon)^2- 4 y \mfp^2}}} \rd y = -\left(\log(\epsilon) - \log(\mfp^2) +2\gamma_E \right) g_n(\mfp^2)\\
&\qquad \qquad - \int_{c-\ri \infty}^{c+ \ri \infty} \mfp^s\left( \psi(-s) + \psi(s+1) \right)\CM\left[ g_n(y); -s \right] {\rd s 
\over 2 \pi \ri}+\CO(\epsilon). 
\ea
\ee
We note that the divergent part, proportional to $\log(\epsilon)$, leads to the result in (\ref{div-fin}). 
The inverse Mellin transform in the second line can be computed in various ways. 
One possibility is to do a sum over residues, 
\be
\ba 
&- \int_{c-\ri \infty}^{c+ \ri \infty} \mfp^{2 s}\left( \psi(-s) + \psi(s+1) \right)\CM\left[ g_n(y); -s \right] {\rd s 
\over 2 \pi \ri}\\
&\qquad \qquad =\sum_{m \ge 0} {\rm Res}_{s=m}  \left( \mfp^{2 s} \left( \psi(-s) + \psi(s+1) \right)\CM\left[ g_n(y); -s \right] \right). 
\ea
\ee
By using this method, one can calculate the functions $\CJ_n(\mfp^2)$ easily for low values of $n$. One finds, for example, 
\be
\CJ_0(\mfp^2)=   -2 {\log(1+ \mfp^2) \over \mfp^2+1}, 
\ee
while for $n=1$ we obtain 
\be
\label{J1}
\CJ_1(\mfp^2)=\frac{12\mfp^2 \rm{Li}_2(-\mfp^2)+\mfp^2 \left(3 (\log (\mfp^2)-4) \log (\mfp^2)+\pi ^2\right)+12 (\mfp^2+\mfp^2 \log (\mfp^2)+1) \log
   (\mfp^2+1)}{6 (\mfp^2+1)^2}. \ee

  It is however possible to obtain a very useful, closed integral formula by using general properties of the Mellin transform. 
  These are 
\be
\CM\left[ -x f'(x);s \right]=s \CM \left[f(x); s\right],
\ee
and the following convolution formula: if $F_i(s)= \CM\left[ f_i(t) ;s \right]$, $i=1,2$, then, 
\be
\CM\left[\int_0^\infty f_1\left( {x \over t} \right) f_2(t) {\rd t \over t} ; s \right]= F_1(s) F_2(s). 
\ee
We also note that 
\be
\psi(s)+ \psi(1-s)= 2 \psi(s)+ \pi \cot(\pi s), 
\ee
as well as the inverse Mellin transforms \cite{mellin}
\be
\ba
\CM^{-1} \left[ \pi \cot(\pi s); x\right]&={\rm P} {1\over 1-x}, \\ 
\CM^{-1} \left[ {\psi(s) \over s}; x\right]&= \begin{cases} -\gamma_E- \log\left( {1\over x}-1 \right), & \text{if $x<1$}\\
0, & \text{if $x>1$} . 
\end{cases}
\ea
\ee
One obtains in this way, 
\be
\label{pvalint}
\ba
&\int_{c-\ri \infty}^{c+ \ri \infty} \mfp^{-2 s} \left( \psi(s) + \psi(1-s) \right)\CM\left[ g_n(y); s \right] {\rd s 
\over 2 \pi \ri}\\
& = {\rm P} \int_0^\infty {g_n(t) \over t-\mfp^2}  \rd t +2  \int_{\mfp^2}^\infty \log\left( {t\over \mfp^2}-1 \right) g_n'(t) \rd t + 2 \gamma_E g_n(\mfp^2), 
\ea
\ee
It follows that, for $n\ge 1$, the functions defined in (\ref{div-fin}) are given by
\be
\label{J-ints}
\CJ_n(\mfp^2)= -\int_0^\infty \left({{\rm P} \over t-\mfp^2} +{1  \over t} \right) g_n(t) \rd t -2  \int_{\mfp^2}^\infty \log\left( {t\over \mfp^2}-1 \right) g_n'(t) \rd t + \log(\mfp^2) g_n(\mfp^2). 
\ee

\subsection{Trans-series for the ground state energy}
\label{app-ts}
We now give some details of the calculations in section \ref{subsec-ts}. 

In order to derive (\ref{eps-def}), we need the explicit integral of the last term in (\ref{Iint}), which can be calculated 
in terms of Bessel, hypergeometric and Meijer functions as follows: 
\be
\label{spec-int}
\ba
- \int_\epsilon^\infty \CR\left({\nu \over z} \right){\rd \nu \over \nu^2}&={\pi \over 2 \epsilon} Y_0\left(\frac{2 \epsilon }{z}\right) \, _1F_2\left(1;-\frac{1}{2},\frac{1}{2};-\frac{\epsilon
   ^2}{z^2}\right)-{\pi \over z} Y_1\left(\frac{2 \epsilon }{z}\right) \,
   _1F_2\left(1;\frac{1}{2},\frac{1}{2};-\frac{\epsilon ^2}{z^2}\right)\\
   & -{\pi \over 4 z } G_{2,4}^{2,1}\left(\frac{\epsilon ^2}{z^2}\biggr|
\begin{array}{c}
 0,1 \\
 0,0,-\frac{1}{2},-\frac{1}{2} \\
\end{array}
\right)\\&=c_0(\epsilon, \sigma) + c_1(\epsilon, \sigma) z^{-1} + \CO(\epsilon), 
\ea
\ee
where 
\be
\label{ceps}
\ba
c_0(\epsilon, \sigma)&=\int_{\epsilon}^\infty {\gamma_E + \log(\nu) - \sigma \over \nu^2} \rd \nu, \\
c_1(\epsilon, \sigma)&=2\left(-2+ \gamma_E + \log(4 \epsilon)- \sigma \right). 
\ea
\ee

In order to verify (\ref{asym-proof}), one notes that the difference between the integral in the first line of (\ref{ennu}) (with no $\gamma$ factor), 
and $ \mathfrak{J}(\sigma_0)$, is given by the limit as $\epsilon \to 0$ of 
\be
\label{spec-int2}
\ba 
 \re^{-\sigma_0 \epsilon} \epsilon^{\epsilon-1} \Gamma(-\epsilon)+{1\over \epsilon^2}+\int_{\epsilon}^\infty \left[  
{\gamma_E + \log(\nu)- \sigma \over \nu^2}-\re^{-{\nu \over \gamma}} \left( {1\over \nu^2} + {\gamma_E + \log(\nu)-\log(\gamma) \over \nu} \right) 
 \right] \rd \nu. 
\ea
\ee
This can be computed e.g. by explicit evaluation of the integrals, and gives
\be
- {1\over 2 \gamma^2}+{1\over\gamma}-{\pi^2 \over 6}, 
\ee
which establishes (\ref{asym-proof}). 

Finally, we analyze the singularities of the integrands of the functions $\CI_n^\pm (\sigma)$ and verify explicitly the cancellation of 
ambiguities. The singularities are 
simple poles at positive integers. We want to show that these poles cancel in the total sum (\ref{z-ts}). This cancellation occurs 
between different terms in the trans-series, similarly to the cancellation of singularities in \cite{bbk}. 
 Since the building blocks of the integrands are the functions $a_k(\nu)$, $b_k(\nu)$, we first look at their singularity structure. 
 \begin{enumerate}
 \item The function $a_k(\nu)$ has simple poles at 
 \be
 \label{nu-a-poles}
 \nu_\star=-k, -k+2, \cdots, k-2, k, 
 \ee
 with a residue given by 
 \be
 \label{nu-a-res}
 c_{\nu_\star}={(-1)^{k+{k-\nu_\star \over 2}} \over \Gamma \left(1+ {\nu_\star+ k\over 2}\right) \left( {k-\nu_\star \over 2}\right)!}. 
 \ee
 They will multiply a power of $z$ given by $z^{-k}$. 
 \item The function $b_{\tilde k}(\nu)$ has poles at $\nu_\star = - \tilde k +r$, $r \in \IZ_{\ge 0}$, 
 and it will appear multiplied by a power 
 \be
 z^{-\nu_\star- 2 \tilde k}= z^{-r -\tilde k}. 
 \ee
 To obtain the same power of $z$ as in the previous pole, we must have 
 $r+ \tilde k=k$, i.e. the index $\tilde k$ is determined by 
 \be
 \label{second-ser}
 \nu_\star+2 \tilde k=k. 
 \ee
 It is easy to see that the residue of $b_{\tilde k}(\nu)$ at this pole is given by $-c_{\nu_\star}$.  
 \end{enumerate}
 
 Let us now see how singularities cancel. Fix a value of $k$ and a pole of $a_k(\nu)$ at $\nu_*$. The part of the integrand 
 involving $a_k(\nu)$ has a simple pole of the form 
 \be
 k z^{-k} a_k (\nu) \nu^{k-2}= z^{-k} \nu_\star^{k-2} { k c_{\nu_\star} \over \nu- \nu_\star} + \text{regular}. 
 \ee
 At the same time, near the same pole $\nu= \nu_\star$, we have to consider the term involving $b_{\tilde k}(\nu)$ 
 with $\tilde k$ determined by (\ref{second-ser}). The part of the integrand involving $b_k (\nu)$ has a simple pole of the form 
 \be
 -z^{-k} \nu_\star^{2 \tilde k + \nu_\star-2}(\nu_\star + 2 \tilde k) { c_{\nu_\star} \over \nu- \nu_\star} + \text{regular}, 
 \ee
 which cancels precisely the previous singularity due to (\ref{second-ser}).

\bibliographystyle{JHEP}

\linespread{0.6}
\bibliography{biblio-renormalons}

\end{document}